\documentclass[twocolumn,showpacs,floats,floatfix,superscriptaddress,aps,pra]{revtex4-1}
\usepackage{amsfonts,amssymb,amsmath}
\usepackage{color,calc}
\usepackage[dvips]{graphicx}
\usepackage{bm}
\usepackage[normalem]{ulem}

\def\sech{\,\text{sech}\,}

\def\e{\,\text{e}}

\def\e{e}

\def\to{\rightarrow}

\def\sech{\text{sech}}

\def\T{\mathcal{T}} 

\def\red{}
\def\black{}

\def\to{\rightarrow}

\begin{document}

\author{Genko T. Genov}
\affiliation{Institut for Quantum Optics, Ulm University, Albert-Einstein-Allee 11, Ulm 89081, Germany}

\author{Yachel Ben-Shalom}
\affiliation{Department of Applied Physics, Rachel and Selim School of Engineering, Hebrew University, Jerusalem 9190401, Israel}

\author{Fedor Jelezko}
\affiliation{Institut for Quantum Optics, Ulm University, Albert-Einstein-Allee 11, Ulm 89081, Germany}

\author{Alex Retzker}
\affiliation{Racah Institute of Physics, The Hebrew University of Jerusalem, Jerusalem 91904, Givat Ram, Israel}

\author{Nir Bar-Gill}
\affiliation{Department of Applied Physics, Rachel and Selim School of Engineering, Hebrew University, Jerusalem 9190401, Israel}
\affiliation{Racah Institute of Physics, The Hebrew University of Jerusalem, Jerusalem 91904, Givat Ram, Israel}

\title{Efficient and robust signal sensing by sequences of adiabatic chirped pulses}

\date{\today}

\begin{abstract}
We propose a scheme for sensing of an oscillating field in systems with large inhomogeneous broadening and driving field variation by applying sequences of phased, adiabatic, chirped pulses. The latter act as a double filter for dynamical decoupling, where the adiabatic changes of the mixing angle during the pulses rectify the signal and partially remove frequency noise. The sudden changes between the pulses act as instantaneous $\pi$ pulses in the adiabatic basis for additional noise suppresion. We also use the pulses' phases to correct for other errors, e.g., due to non-adiabatic couplings. Our technique improves significantly the coherence time
%
%
in comparison to standard XY8 dynamical decoupling in realistic simulations in NV centers with large inhomogeneous broadening and is suitable for experimental implementations with substantial driving field inhomogeneity. 
Beyond the theoretical proposal, we also present proof-of-principle experimental results for quantum sensing of an oscillating field in NV centers in diamond, demonstrating superior performance compared to the standard technique.
\end{abstract}

\maketitle

\emph{Introduction.---}\label{Section:Introduction}
%
Magnetometry experiments require the measurement of a signal whose characteristics are related to a magnetic field to be sensed.
Pulsed and continuous dynamical decoupling have already been applied for quantum memories and for sensing of oscillating (AC) fields in various systems, e.g., trapped ions, nitrogen-vacancy (NV) centers in diamond, rare-earth doped-solids  \cite{Viola09PRL,Suter16RMP,WrachrupNano2013,DegenARPC2014,DegenRMP2017,BalasubramanianNatMat2009,deLangeScience2010,WrachtrupPRB2011,
KnowlesNatMat2014,HollenbergNatNano2011,WalsworthNature2013,LukinNature2013,BalasubramanianOpinBio2014,TimoneyNature2011,HirosePRA2012,
AielloNatComm2013,Heinze13PRL,Schmitt2017Science,StarkNatComm2017,StarkSciRep2018,DegenRMP2017,HirosePRA2012,AielloNatComm2013,StarkNatComm2017,
StarkSciRep2018,Sriarunothai2019QST,GenovPRL2017,RDD_review12Suter,CaiNJP2012,CohenFP2017,AharonPRL2019,
FarfurnikJOpt2018,FarfurnikPRB2015,CasanovaPRA2015,JoasNatComm2017,ZhangJPhysDApplPhys2018,GenovQST2019}. However, the sensitivity is reduced in systems with large inhomogeneous broadening
and field inhomogeneities also limit efficiency.
Then, only a small fraction of the sensor atoms contribute to the signal due to the limited bandwidth of the control field.

Adiabatic chirped pulses perform robust population flips by rapid adiabatic passage (RAP) even with inhomogeneous broadening, a weak driving field, and significant amplitude fluctuations \cite{Shore1990Book,Vitanov01ARPC,Tannus97AdiabBook,GarwoodJMR1991,ZlatanovPRA2020}. They have been applied for rephasing of atomic coherences  \cite{Lauro11PRA,Mieth12PRA,Pascual-Winter13NJP} and combined with composite pulses \cite{Levitt97Review,Tycko84,Torosov11PRA} for high fidelity population transfer \cite{Torosov11PRL,Schraft13PRA,Genov2014PRL,Genov2018PRA}.

\begin{figure}[t!]
\includegraphics[width=\columnwidth]{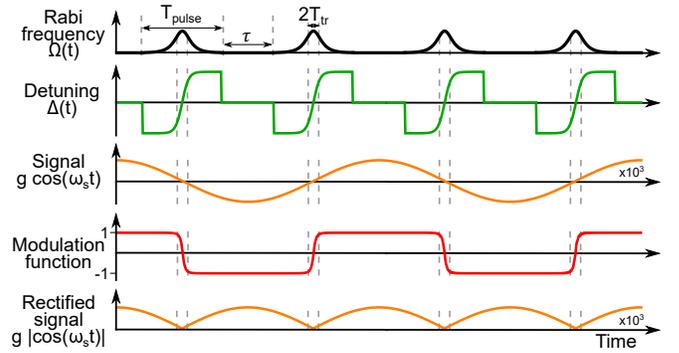}
\caption{(color online)
Scheme for AC magnetometry with RAP pulses. The example follows the Allen-Eberly (AE) model \cite{Allen-Eberly1987} with $\Omega(t)=\Omega_0 \sech{\left[(t-t_{c,k})/T\right]}$, where $t_{c,k}$ is the k-th pulse center, $\Delta(t)=(R/2)\tanh{\left[(t-t_{c,k})/T\right]}$, $R$ is the chirp range, $T_{\text{pulse}}=12T$.
The pulses can be phase-shifted 
for improved performance. Our goal is to sense  
an oscillating signal with $\omega_{s}=\pi/(T_{\text{pulse}}+\tau)$. The RAP pulses perform population transfer and modulate the signal. Usually, it is advantageous to take $\tau=0$ and increase $T_{\text{pulse}}$ to match $\omega_{s}$, keeping $T$ constant, as this improves adiabaticity without affecting the modulation function. The transition time $T_{\text{tr}}=1/|\nu^{\prime}(t_{c,k})|=4T\Omega_0/R$ characterizes the time scale of population transfer. 
When $T_{\text{tr}}\ll T+\tau$, its effect can be neglected and the rectified signal takes the bottom shape. }
\label{Fig1:RAP_scheme}
\end{figure}

In this letter we propose sequences of phased RAP pulses for dynamical decoupling (DD) and sensing of an AC field. The signal has a frequency of half the pulses' repetition rate and can be sensed in systems with large field inhomogeneity and varying transition frequencies, e.g., due to inhomogeneous broadening or different atom orientations with respect to the quantization axis as with NV containing nanodiamonds in cells.
The RAP sequences act as a double filter for DD, where the population transfer during a pulse rectifies the signal and partially removes frequency noise. The sudden changes in the mixing angle between the pulses act as fast $\pi$ pulses in the adiabatic basis for additional noise compensation. 
Finally, we use the pulses' phases as control parameters to correct for other errors, e.g., due to non-adiabatic couplings. We demonstrate the superior performance
of the RAP protocol with the XY8 sequence (RAP-XY8) in comparison to the widely used XY8 sequence \cite{DegenRMP2017} with rectangular pulses with the same peak Rabi frequency for realistic simulations.
Finally, we present data from a demonstration quantum sensing experiment in an ensemble of NV centers in diamond. As the
simulations and experimental data show,
RAP sensing significantly outperforms the standard protocol.

%
%


\begin{figure*}[t!]
\includegraphics[width=\textwidth]{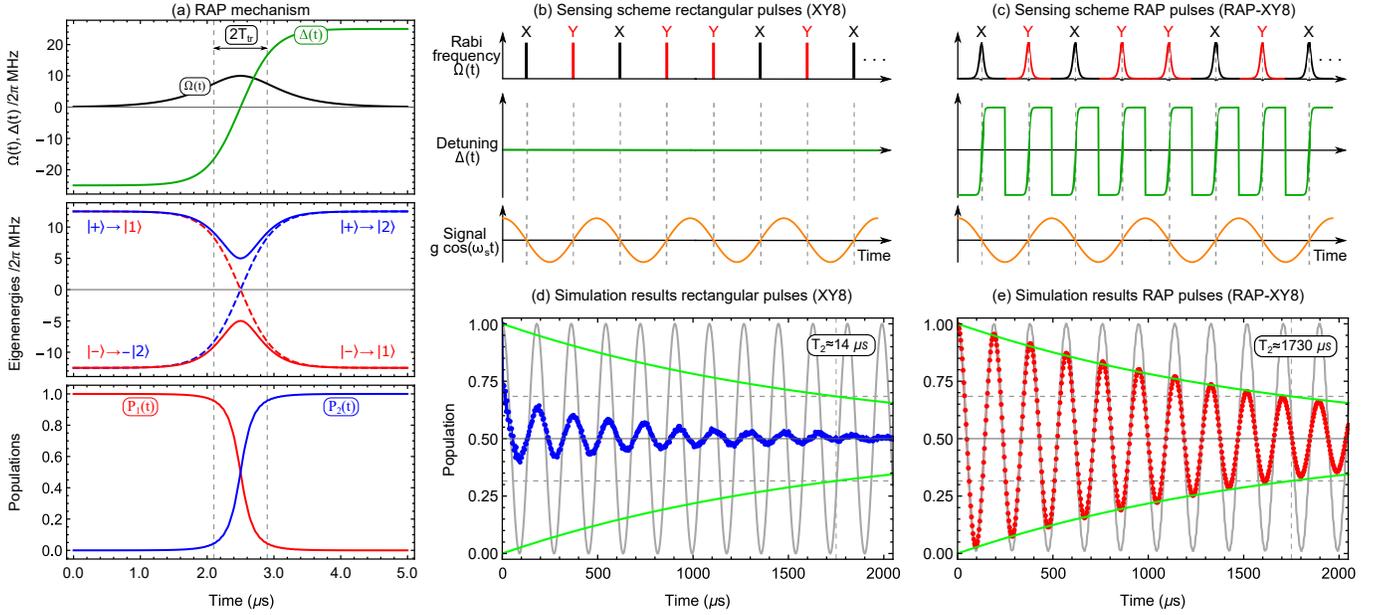}
\caption{(color online)
(a) RAP mechanism: (\emph{top}) Rabi frequency and detuning for the AE model with 
$\Omega_0=2\pi~10$ MHz, 
chirp range $R=2\pi~50$ MHz, and $T_{\text{pulse}}=10T=5~\mu$s. (\emph{middle}) Eigenenergies in the bare (dashed lines) and adiabatic (solid lines) bases. The composition of the adiabatic states changes, leading to population transfer in the bare basis due to a level crossing. (\emph{bottom}) Simulation of population transfer, characterized by the transition time $T_{\text{tr}}=1/|\nu^{\prime}(t_{c})|=4T\Omega_0/R=0.4~\mu$s.
Scheme for sensing with (b) standard XY8 and (c) RAP-XY8. In both, the atoms are prepared initially in $|1_{y}\rangle$.
Corresponding numerical simulations of the population in the $|1_{y}\rangle$, observed directly in the bare basis at time intervals of $8~
\mu$s with DD by (d) XY8 with rectangular pulses with $\Omega(t)=\Omega_0=2\pi~10$ MHz, duration $T_{\text{pulse}}=50$ ns and pulse separation $\tau=0.95~\mu$s, and (e) RAP-XY8
with $\Omega_0=2\pi~10$ MHz, a target chirp range $R=2\pi~95$ MHz, characteristic time $T=0.2~\mu$s, pulse duration $T_{\text{pulse}}=1~\mu$s, and pulse separation $
\tau=0$. The sensed field has an amplitude of $g=2\pi~4.34$ kHz, initial phase $\xi=0$ and angular frequency $\omega_{s}=2\pi~0.5$ MHz. The peak Rabi frequency is the same in both protocols and $T_{\text{pulse}}+
\tau=\pi/\omega_{s}$. 
The slight delay in the ideal, theoretical gray curve with RAP from $p=\cos(\eta(t))^2$ is mainly to the non-instantaneous transition time, which is taken into account in the simulation.
}
\label{Fig:rect_RAP_comparison}
\end{figure*}

\emph{Theory of RAP sensing.---}\label{Section:RAP_sensing}
We consider a two-state system, described by the Hamiltonian in the rotating-wave approximation
\begin{align}
H_{\text{s}}(t)&=-\frac{\widetilde{\Delta}(t)}{2}\sigma_{z}+\frac{\widetilde{\Omega}(t)}{2}\sigma_{x}+g\sigma_{z}\cos{(\omega_{\text{s}}t+\xi)},
\end{align}
where $\widetilde{\Delta}(t)\equiv \Delta(t)-\Delta_{\epsilon}(t)$ is the detuning, 
which depends on the target detuning $\Delta(t)$ 
and an error $\Delta_{\epsilon}(t)$. 
The Rabi frequency is
$\widetilde{\Omega}(t)=\Omega(t)[1+\epsilon_{\Omega}(t)]=\mathbf{\mu} \widetilde{B}(t)$, where
$\epsilon_{\Omega}(t)$ is an error term. 
The amplitude, angular frequency, and initial phase of the sensed AC field are $g$, $\omega_{\text{s}}$, and $\xi$.
The Hamiltonian in the adiabatic basis is \cite{Shore1990Book,Vitanov01ARPC}
\begin{align}\label{Eq:H_ad_sense}
H_{\text{ad,s}}(t)&= -\frac{\widetilde{\Omega}_{\text{eff}}(t)}{2}\sigma_{z}
+g\cos{(\omega_{\text{s}}t+\xi)}\\
&\times\left[\cos{(2\widetilde{\nu}(t))}\sigma_{z}+\sin{(2\widetilde{\nu}(t))}\sigma_{x}\right],\notag
\end{align}
where
$\widetilde{\nu}(t)=\arctan{\left[-\frac{\widetilde{\Delta}(t)}{\widetilde{\Omega}(t)}+\sqrt{1+\frac{\widetilde{\Delta}(t)^2}{\widetilde{\Omega}(t)^2}}\right]}$
is the mixing angle \cite{Shore1990Book} (see Appendix \ref{Section:RAP_Theory}), $\widetilde{\Omega}_{\text{eff}}(t)=\sqrt{\widetilde{\Omega}(t)^2+\widetilde{\Delta}(t)^2}$, and we applied the adiabatic approximation ($|\widetilde{\nu}^{\prime}(t)|\ll\widetilde{\Omega}_{\text{eff}}(t)$).
One can obtain intuition about
the effect of a RAP pulse
by considering the adiabatic and bare bases (see Fig. \ref{Fig:rect_RAP_comparison}(a) and Appendix \ref{Section:RAP_Theory}).
Adiabatic evolution with population inversion is not necessarily optimal for DD due to noise in $\widetilde{\Omega}_{\text{eff}}(t)$ (see Appendix \ref{Section:Robust_RAP_sequences}).
Thus, we consider sequences of RAP pulses where the detuning shifts between the pulses, e.g., from large positive to large negative values, resulting in shifts of the mixing angle by $\Delta\nu\approx\pi/2$. This is equivalent to applying instantaneous $\pi$ pulses in the adiabatic basis, which compensate the noise in $\widetilde{\Omega}_{\text{eff}}(t)$.
The time scale of the shifts is limited by the sampling rate of the microwave control synthesizer, e.g. an arbitrary waveform generator, so sub-nanosecond shifts are readily achievable.

We incorporate such changes in the definition of our interaction basis,
and obtain the Hamiltonian (see Appendix \ref{Section:RAP_sensing})
\begin{equation}\label{Eq:H_ad_sense_interaction_tog_approx}
H_{\text{int,tog,s}}(t)= -\widetilde{f}(t)g\cos{(\omega_{\text{s}}t+\xi)}\sigma_{z},
\end{equation}
where the modulation function $\widetilde{f}(t)= f(t)\cos{(2\widetilde{\nu}(t))}$ with $f(t)=-1$ ($f(t)=1$) during the odd (even) pulses. 
We neglected fast oscillating terms, assuming $\omega_{\text{s}}\ll \widetilde{\Omega}_{\text{eff}}(t)$ and $|\widetilde{\nu}^{\prime}(t)|\ll\widetilde{\Omega}_{\text{eff}}(t)$.
We note that $\widetilde{f}(t)$ stays the same if 
$\Delta\nu=\pm\pi/2$ between two RAP pulses because $f(t)$ and $\cos{(2\widetilde{\nu}(t))}$ change their signs simultaneously. Thus, the $\widetilde{f}(t)$ is affected only by adiabatic changes in the mixing angle (see Fig. \ref{Fig1:RAP_scheme}).
As the fast shifts do not affect $\widetilde{f}(t)$ (but only $f(t)$), the pulses can be truncated and separated by free evolution time $\tau$ (see Fig. \ref{Fig1:RAP_scheme}), and we can sense the signal if  $T_{\text{pulse}}+\tau=\pi/\omega_{\text{s}}$.
However, it is usually preferable to use long pulses and $\tau=0$ as this improves adiabaticity. 
%
If the RAP transition time is short, i.e., $T_{\text{tr}}\ll \pi/\omega_{\text{s}}$, where $T_{\text{tr}}=2\widetilde{\Omega}(t_{c})/\widetilde{\Delta}^{\prime}(t_{c})$ is defined in analogy to the transition time in stimulated Raman adiabatic passage \cite{Boradjiev2010PRA}, and $t_{c}$ is the time of level crossing in the bare basis, the modulation function is approximately a step function (see Fig. \ref{Fig1:RAP_scheme} and Appendix \ref{Subsection:RAP_transition_time}).
Then, the Hamiltonian in Eq. \eqref{Eq:H_ad_sense_interaction_tog_approx} becomes
$H_{\text{int,tog,s}}(t)\approx -g|\cos{(\omega_{\text{s}}t)}|\sigma_{z}$ where we assumed $\xi=0$ for maximum contrast and $\nu(t_0)=\pi/2$. 
Similarly to standard pulsed DD, the sensor qubit performs Ramsey oscillations 
and accumulates a phase $2\eta(t)$, 
where ($g\ll \omega_{\text{s}}$): $\eta(t)\equiv\int_{0}^{t}g|\cos{(\omega_{\text{s}} t^{\prime})}| d t^{\prime}\approx\frac{2}{\pi}g t.$
%
We can observe the signal stroboscopically in the bare basis after every second RAP pulse when the dynamic phase due to $\widetilde{\Omega}_{\text{eff}}(t)$ (and its noise) is compensated by the instantaneous changes in the mixing angle. 
We note that these fast shifts do not compensate non-adiabatic couplings, which 
usually commute with the Hamiltonian during the shifts. Additionally, $\Delta\nu$ might differ from $\pi/2$. 
We use the relative phases of the pulses to compensate for such and other errors (see Appendix \ref{Subsection:Phased_RAP}), e.g., using the popular XY, KDD, or UR sequences \cite{RDD_review12Suter,Torosov11PRL,Genov2014PRL,GenovPRL2017}. These are based on composite pulses and improve population transfer and rephasing \cite{Torosov11PRL,Schraft13PRA,GenovPRL2017,Genov2018PRA}.


%
We compare the performance of standard rectangular and RAP pulses by a numerical simulation for DD in a two-state system subject to magnetic noise and driving field fluctuations that follow an Ornstein-Uhlenbeck process \cite{UhlenbeckRMP1945,GillespieAJP1996} and are typical for experiments in NV centers \cite{CaiNJP2012,AharonNJP2016}. We also assume an inhomogeneous broadening, leading to 
dephasing time of $T_{2}^{\ast}\approx 20$ ns and a Hahn echo $T_{2}\approx 13~\mu$s (see Appendix \ref{Section:Numerics}).
We choose the Allen-Eberly model for RAP amplitude and detuning   \cite{HioePRA1984,Allen-Eberly1987,Kyoseva06PRA} due to its preferable adiabaticity (see Appendix \ref{Subsection:AE}). 

Figure \ref{Fig:rect_RAP_comparison}(b-e) shows the scheme and simulated evolution of the population in the $|1_{y}\rangle$ state in the bare basis for sensing with rectangular and RAP pulses with identical peak Rabi frequency and the phases of the widely used XY8 sequence.
Due to the inhomogeneous broadening the contrast is lost quickly with the standard XY8, which has a $T_2\approx 14~\mu$s and is increased by more than two orders of magnitude to $T_2\approx 1.7$ ms with RAP-XY8.
The remaining decay for RAP-XY8 is mainly due to high frequency components of the noise 
and imperfect adiabaticity. The coherence time with RAP-XY8 approaches the population lifetime of an NV center, which can reach up to $6$ ms \cite{Bar-GillNatComm2013} and is not considered in the simulation.

\begin{figure*}[t!]
\includegraphics[width=\textwidth]{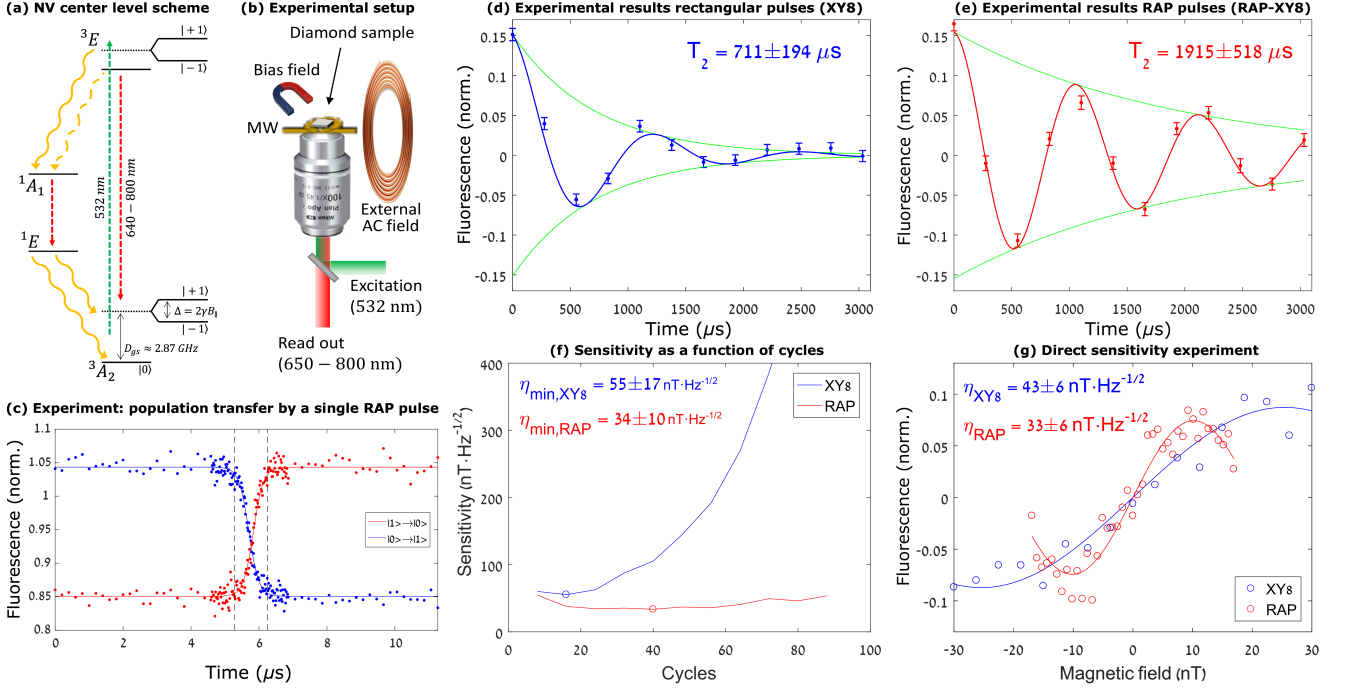}
\caption{(color online)
(a) Energy level diagram of the NV center. The NV is excited with off-resonant green light (532 nm), coherently manipulated using microwave fields, and optically read-out through spin-dependent fluorescence between states \(\left|0\right\rangle\) and \(\left|\pm1\right\rangle\). (b) Experimental setup scheme. (c) Measurement of population transfer by a RAP pulse with \(\Omega_{0}= 2\pi\:5\:\) MHz, duration \(T_{\text{pulse}}=11.4\:\mu s\), a target chirp range \(R=2\pi\:40\: \) MHz, and characteristic time \(T/T_{\text{pulse}}=0.17\).
The transition time is short compared to the pulse duration ($T_{tr}\approx 0.07 T_{\text{pulse}}$), as expected from theory. The time between the dashed lines depicts the theoretical transition time \(4\Omega_{0}T/R\) (see Appendix, sec).
Coherence decay curves in presence of an artificial external magnetic field with amplitude of \(52\:nT\) and frequency of \(14.5\:\)kHz for the low Rabi frequency experiment \(\Omega_{0}=2\pi\:1.7\:\) MHz (see Appendix, sec. \ref{Section:SM_Experimental_results}) for (d) XY8 and (e) for RAP-XY8. (f) Sensitivity vs. number of free evolution-pulse-free evolution cycles. (g) Variation of fluorescence vs. amplitude of the sensed magnetic field for a direct measurement of the sensitivity at its optimal points, marked with a circle in (f), i.e., for 16 (40) cycles for XY8 (RAP-XY8). The steeper curve slope with RAP-XY8 shows improved sensitivity; the slightly lower contrast with RAP-XY8 is due to the higher pulse number.
}
\label{fig:low Rabi details}
\end{figure*}

\emph{Experimental demonstration.---}\label{Section:Exp_description}
%
We experimentally demonstrate our scheme in an ensemble of NV centers in diamond. The measurements are performed in a home-built confocal fluorescence microscope at a magnetic field of 332 Gauss. The system and experimental setup are described in Fig. \ref{fig:low Rabi details}(a-b) and in \cite{MeirzadaPRB2018}.
%
%
Modulated microwave control fields are created using an arbitrary waveform generator (AWG - Tektronix AWG70002A - 16 Gs). The measurements are performed on a standard-grade diamond sample (Element Six) with an NV density of $\sim 10$ ppb. The NV spin properties are measured $T_{1}=5.8 \pm 0.6 $ms, $T_{2}^{\star}=34 \pm 14$ ns, Hahn echo $T_{2}=198 \pm 18 \mu s$, inhomogeneous broadening of $ 2 \pi \times (2.1 \pm 0.1)$ MHz. The NV is initially prepared in state $|0\rangle$ by optical pumping.

We carried out three sets of experiments at
(i) high Rabi frequency ($2 \pi~5$ MHz), (ii) high Rabi frequency with artificially added amplitude errors with a Gaussian distribution with a width of \(0.2~\Omega_{0}\), (iii) low Rabi frequency ($2 \pi~1.7$ MHz), comparable to the bandwidth of the inhomogeneous broadening.
Each set included regular XY8 (rectangular pulses) and RAP XY8. We performed two measurements for each data point with a rectangular $\pi/2$ (half RAP) $X$ pulse for initialization and $\pm X$ before optical readout that match the corresponding parameters of the XY8 (RAP-XY8) DD pulses. We recorded the fluorescence difference between the two measurements. 
The RAP chirp range and characteristic time were optimized experimentally (see Appendix \ref{Section:SM_Experimental_results}).




%
First, we measured the NV electron spin coherence time with DD. 
RAP-XY8 achieved $T_2 \simeq 1943~\mu$s for all experiments and was robust to amplitude noise and inhomogeneous broadening (see Appendix \ref{Section:SM_Experimental_results}). The coherence time for the standard XY8 had $T_2 \simeq 1811~\mu$s only for the optimized high-Rabi experiment but it degraded by about 40\% with amplitude noise ($T_2 \simeq 1192~\mu$s) and lower Rabi frequency ($T_2 \simeq 1057~\mu$s). As expected from theory, RAP-XY8 performed better for the non-ideal cases. The coherence times of standard XY8 is expectedly longer than in the simulation in Fig. \ref{Fig:rect_RAP_comparison} due to the lower inhomogeneous broadening in the experiment.


Next, we performed quantum sensing by adding an external magnetic field to characterize the magnetic sensitivity under RAP and regular control pulses. The AC magnetic signal was generated using a home-built coil, driven by a function generator (Rigol 5252).
The presence of the external field reduced the coherence time by $\sim 30 \%$ using standard XY8 (see Fig. \ref{fig:low Rabi details}(d)). We attribute this effect to the rotation of the NV electron spin state in the XY plane of the Bloch sphere due to the field, so the system is more sensitive to pulse imperfections. While this effect should be negligible in the small-field limit \cite{PhamPRB2012}, it does not adversely affect RAP even for larger fields (Fig. \ref{fig:low Rabi details}(e)).

Next, we analyzed the sensitivity to an AC field, defined as \(\eta=\frac{\sigma}{\partial S/\partial B}\sqrt{T_m}\) \cite{PhamPRB2012}, where $\sigma$ is the standard deviation of the single point fluorescence data, \(\partial S/\partial B\) is the maximal slope in the curve of fluorescence signal vs. magnetic field, and \(T_m\) is the time of a single measurement. Figure \ref{fig:low Rabi details}(f) shows the robust sensitivity of RAP-XY8 as a function of the number of pulses when extracted indirectly from the data. Direct measurements of the sensitivity in Fig. \ref{fig:low Rabi details}(g) under optimized conditions exhibit a $\sim 30 \%$ improvement for RAP sequences over regular XY8. 
%
Standard XY8 is affected by pulse errors, so sensitivity is strongly suppressed when we apply many puless. In contrast, RAP-XY8 remains robust 
with a nearly constant value over a large range of applied pulses. This indicates that RAP allows for sensing of weaker fields and improved precision due to the longer achievable coherence times.
We note that in our experiments the inhomogeneous broadening was relatively low (approx. $\simeq 2$ MHz), and thus we simulated a larger broadening by applying either weak driving (low Rabi frequency of $\simeq 1.7$ MHz) or added driving noise. The parameters still correspond to limited broadening, yet clearly demonstrate a significant improvement in both coherence time and magnetic sensitivity of RAP compared to standard sensing.


\black

\emph{Discussion.---}\label{Section:Discussion}
%
%
%
RAP improved performance is due to its broad bandwidth and robustness.
%
%
%
Its transition probability error depends on the pulse shape \cite{Vitanov01ARPC} and can be estimated 
\cite{LandauZener1932} (see Appendix \ref{Section:RAP_simulation_comparison}) 
%
%
$\epsilon_{\text{RAP}}\sim\epsilon_{\text{rect}}\frac{\omega_{s}^2}{\Omega_0^2}$ where $\epsilon_{\text{rect}}\sim\Delta_{\text{inh}}^2/\Omega_{0}^2$ is the rectangular pulse error.
Thus, RAP sensing improves performance when $\Omega_{0}<\Delta_{\text{inh}}$ and $\omega_{s}\ll\Omega_{0}$. It is also less sensitive to amplitude variation. 
%
%
%
The frequency range of RAP sensing can be estimated as
$
\pi \left(\frac{b^2}{12\widetilde{\tau}}\right)^{1/3}\ll\omega_{s}\ll \frac{\pi^2\Omega_{0}^2}{4\Delta_{\text{inh}}},
$
(see Appendix \ref{Section:RAP_simulation_comparison}) where the lower limit depends on the homogeneous broadening noise spectrum 
$S(\omega)=\frac{b^2}{\pi}\frac{1/\widetilde{\tau}}{(1/\widetilde{\tau})^2+\omega^2}$, where $\widetilde{\tau}$ is the correlation time of the environment and $b$ 
is the bath coupling strength.
%
The upper limit can increase significantly by using appropriate pulse shapes or phased sequences, e.g., XY8. 

When the the RAP transition time condition $T_{\text{tr}}\ll\omega_s/\pi$ is not fulfilled, there can be a slight shift in the amplitude of the detected AC field in the noiseless case (see Fig. \ref{Fig:rect_RAP_comparison}).
Imperfect preparation and readout by standard $\pi/2$ pulses can be improved by, e.g., using adiabatic half passage (see Appendix \ref{Section:Preparation}), 
robust composite $\pi/2$ pulses \cite{Levitt97Review}, adiabatic robust pulses \cite{Tannus97AdiabBook,GarwoodJMR1991,ZlatanovPRA2020}, single-shot shaped pulses \cite{NdongJPB2015,Van-DammePRA2017}, pulses designed by optimal control \cite{SkinnerJMR2003,HaberlePRL2013,BraunNJP2014,ScheuerNJP2014,DoldeNatComm2014,NobauerPRL2015}.
%
%
Finally, we note that other methods
\cite{StarkNatComm2017,StarkSciRep2018,FarfurnikJOpt2018,FarfurnikPRB2015,CasanovaPRA2015,JoasNatComm2017,GenovQST2019}
also allow for significant increase in the coherence times in NV centers in diamond, e.g., by applying strong driving fields or robust phased DD sequences. However, these typically require inhomogeneous broadening or amplitude errors to be smaller than the Rabi frequency, e.g., around ten percent. In contrast, RAP pulses are robust to a much broader range of errors.


\emph{Conclusion.---}\label{Section:Conclusion}
%
We theoretically developed and experimentally demonstrated robust and efficient signal sensing by sequences of phased RAP pulses. The signal has a frequency of half the pulses' repetition rate and the sensor qubits can experience large variation in field amplitudes and transition frequencies, e.g., due to inhomogeneous broadening.
%
%
We showed that RAP-XY8 significantly outperforms the standard XY8 protocol with rectangular pulses in a realistic simulation. 
We also performed a proof-of-principle experimental demonstration of RAP sensing in ensembles of NV centers and demonstrated its advantages for systems with inhomogeneous broadening and imperfect, non-uniform microwave driving fields.
The robustness and flexibility of the technique make it applicable for a wide range of experimental platforms, e.g. NV ensembles, NVs in nanodiamonds in living cells, rare-earth doped solids, trapped ions.

\acknowledgments

G. G. and Y. B.-S. contributed equally to this work.
We acknowledge useful discussions with Philipp Neumann and Jochen Scheuer (NVision Imaging Technologies). G. G. acknowledges support of the European Union under grant agreement No. 667192-Hyperdiamond
under the Horizon 2020 program. F. J. acknowledges the support of ERC, BMBF, DFG, Landesstiftung BW, and VW Stiftung. A. R. acknowledges the support of ERC grant QRES, project No. 770929, grant agreement No. 667192-Hyperdiamond under the Horizon 2020 program, the MicroQC, the ASTERIQS and the DiaPol projects. N.B. acknowledges support from the European Union’s Horizon 2020 research and innovation program under grant agreements No. 714005 (ERC StG Q-DIM-SIM), No. 820374 (MetaboliQs) and No. 828946 (PATHOS), and has been supported in part by the Minerva ARCHES award, the CIFAR-Azrieli global scholars program and the Ministry of Science and Technology, Israel.

\appendix

\section{Detailed theory of rapid adiabatic passage}\label{Section:RAP_Theory}

\subsection{The System}

We provide a description of rapid adiabatic passage (RAP) in this section. A detailed review can be found in \cite{Vitanov01ARPC}.
We consider a two-state quantum system with an (angular) transition frequency $\widetilde{\omega}_0(t)$ subject to a control field with a time-dependent carrier frequency $\omega(t)$, where we have assumed that the transition frequency $\widetilde{\omega}_0(t)=\omega_0+\Delta_{\epsilon}(t)$ might vary by $\Delta_{\epsilon}(t)$ from its expected value $\omega_0$, e.g., due to inhomogeneous broadening or magnetic field fluctuations. The evolution of the system without a sensed field is governed by the Hamiltonian ($\hbar=1$)
\begin{equation}\label{Eq:H_general}
\widetilde{H}(t)=\frac{\widetilde{\omega}_0(t)}{2}\sigma_{z}+\widetilde{\Omega}(t)\sigma_{x}\cos{\left(\int_{t_0}^{t}\omega(t^{\prime}) d t^{\prime}+\phi\right)},
\end{equation}
where $\widetilde{\Omega}(t)=\mathbf{\mu} \mathbf{\widetilde{B}}(t)$ is the Rabi frequency, which depends on the dipole moment $\mu$ and the envelope of the applied control field $\widetilde{B}(t)$. The actual Rabi frequency can also be presented as  $\widetilde{\Omega}(t)=\Omega(t)[1+\epsilon_{\Omega}(t)]$, where $\Omega(t)$ is the target Rabi frequency we want to apply and $\epsilon_{\Omega}(t)$ is an error term, e.g., due to amplitude fluctuations and/or inhomogeneity. Additionally, $\phi$ is the initial phase of the control field at the time $t_0$ at the beginning of the interaction, $\sigma_{x}$ and $\sigma_{z}$ are the respective Pauli matrices. Usually only the relative changes of $\phi$ are important.

The angular frequency of the control field can also be presented in terms of its detuning $\Delta(t)$ from the expected transition frequency of the atom as $\omega(t)\equiv\omega_0+\Delta(t)$, so the Hamiltonian takes the form
\begin{equation}\label{Eq:H_general_simplified}
\widetilde{H}(t)=\frac{\omega_0+\Delta_{\epsilon}(t)}{2}\sigma_{z}+\widetilde{\Omega}(t)\sigma_{x}\cos{\left(\omega_0 t +\delta(t)\right)},
\end{equation}
where $\delta(t)\equiv\int_{t_0}^{t}\Delta(t^{\prime}) d t^{\prime}$ is an accumulated phase due to the detuning $\Delta(t)$, $\widetilde{\Omega}(t)\equiv\Omega(t)(1+\epsilon_{\Omega}(t))$
is the actual Rabi frequency of the driving field, and we took $\phi=0$ without loss of generality. 

It is advantageous to move to the rotating frame with respect to $\omega(t)\sigma_{z}/2$ and apply the rotating-wave approximation ($|\widetilde{\Omega}(t)|\ll\omega(t)$) to obtain the Hamiltonian
\begin{align}
H(t)=&-\frac{\widetilde{\Delta}(t)}{2}\sigma_{z}+\frac{\widetilde{\Omega}(t)}{2}\sigma_{x},
\end{align}
where $\widetilde{\Delta}(t)\equiv \Delta(t)-\Delta_{\epsilon}(t)$ is the actual detuning, experienced by a sensor atom, which depends on the detuning $\Delta(t)$ of the driving field from $\omega_0$ and the variation of the actual transition frequency of the atom $\Delta_{\epsilon}(t)$ from $\omega_0$. We will use this Hamiltonian further on in the analysis and will call the quantum states in this basis the bare states.

\begin{figure*}[t!]
\includegraphics[width=1.85\columnwidth]{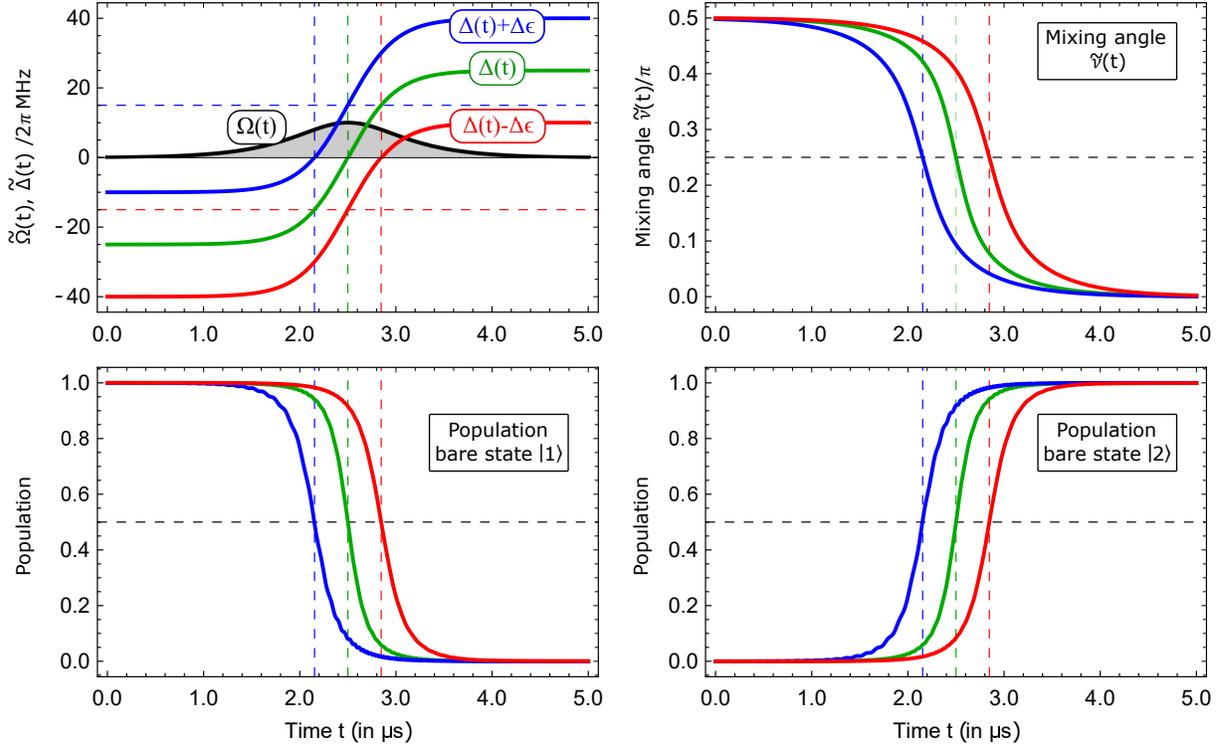}
\caption{(color online)
Numerical simulation of an example for population transfer with RAP for three different detuning errors $\Delta_{\epsilon}(t)=2\pi~\Delta\epsilon$ (blue), $\Delta_{\epsilon}(t)=0$ (green), $\Delta_{\epsilon}(t)=-\Delta\epsilon$ (red) with $\Delta\epsilon=2\pi~15$ MHz. (a) Time-dependence of the respective parameters, used in the simulation: Rabi frequency $\widetilde{\Omega}(t)=\Omega_0 \sech{(t/T)}$, where the peak Rabi frequency $\Omega_0=2\pi~10$ MHz, detuning $\widetilde{\Delta}(t)=\Delta(t)+\Delta_{\epsilon}(t)$ with $\Delta(t)=(R/2)\tanh{(t/T)}$, where $R=2\pi~50$ MHz is the target chirp range, $T=0.5~\mu$s. Time evolution of (b) the mixing angle $\widetilde{\nu}(t)$, (c) the population of the bare state $|1\rangle$, (d) the population of the bare state $|2\rangle$ for the respective values of $\Delta_{\epsilon}(t)$. The population transfer efficiency is very robust to detuning errors even though they are greater than the peak Rabi frequency but the flip of the quantum state happens at different times for each value of $\Delta_{\epsilon}(t)$, i.e., the transfer process is centered at the time when the respective mixing angle is $\widetilde{\nu}(t)=\pi/4$. }
\label{Fig1:RAP_mechanism}
\end{figure*}

\subsection{The adiabatic basis}\label{Subsection:Adiabatic_basis}

It proves useful to consider the evolution of the system in the adiabatic (dressed) basis by making another transformation $\mathbf{d}(t)=R_{\text{ad}}(t)\mathbf{c}(t)$, where $\mathbf{d}(t)=\left[d_{-}(t),d_{+}(t)\right]^{\text{T}}$ are the probability amplitudes of the adiabatic states and $\mathbf{c}(t)=\left[c_{1}(t),c_{2}(t)\right]^{\text{T}}$ are the probability amplitudes of the bare states \cite{Shore1990Book}
\begin{align}\label{Eq:R_matrix_appendix}
R_{\text{ad}}(t)&=\left[\begin{array}{cc} \cos{\widetilde{\nu}(t)}  & -\sin{\widetilde{\nu}(t)}, \\
\sin{\widetilde{\nu}(t)} & \cos{\widetilde{\nu}(t)} \end{array} \right],\\
\widetilde{\nu}(t)&=\arctan{\left[-\frac{\widetilde{\Delta}(t)}{\widetilde{\Omega}(t)}+\sqrt{1+\frac{\widetilde{\Delta}(t)^2}{\widetilde{\Omega}(t)^2}}\right]}.\notag
\end{align}
One can show that $\tan[2\widetilde{\nu}(t)]=\widetilde{\Omega}(t)/\widetilde{\Delta}(t)$, so the mixing angle can in principle also be expressed
by the more intuitive $\widetilde{\nu}(t)=(1/2)\arctan{[\widetilde{\Omega}(t)/\widetilde{\Delta}(t)]}$ \cite{Vitanov01ARPC}. However, the definition in Eq. \eqref{Eq:R_matrix_appendix} is more general as $\arctan{[\widetilde{\Omega}(t)/\widetilde{\Delta}(t)]}$ is defined modulo $\pi$ and it does not allow us to distinguish between $\widetilde{\nu}(t)=0$ and $\pi/2$.
Then, the Hamiltonian in the adiabatic basis becomes
\begin{align}\label{Eq:H_ad_no_approx_appendix}
H_{\text{ad}}(t)&=R_{\text{ad}}(t)H(t)R_{\text{ad}}(t)^{\dagger}-i R_{\text{ad}}(t)\left(\partial_{t}R_{\text{ad}}(t)^{\dagger}\right) \notag\\
H_{\text{ad}}(t)&=\left[\begin{array}{cc} \widetilde{\epsilon}_{-}(t)  & -i\widetilde{\nu}^{\prime}(t)\\
i\widetilde{\nu}^{\prime}(t) & \widetilde{\epsilon}_{+}(t) \end{array} \right],
\end{align}
where $\widetilde{\epsilon}_{\pm}(t)=\pm\frac{1}{2}\sqrt{\widetilde{\Omega}(t)^2+\widetilde{\Delta}(t)^2}$ are the eigenenergies of the adiabatic states and $\widetilde{\nu}^{\prime}(t)$ is the non-adiabatic coupling. When $\widetilde{\nu}(t)$ changes very slowly, i.e.,
\begin{equation}
|\widetilde{\nu}^{\prime}(t)|\ll\widetilde{\epsilon}_{+}(t)-\widetilde{\epsilon}_{-}(t)=\widetilde{\Omega}_{\text{eff}}(t),
\end{equation}
where the effective Rabi frequency $\widetilde{\Omega}_{\text{eff}}(t)\equiv\sqrt{\widetilde{\Omega}(t)^2+\widetilde{\Delta}(t)^2}$, we can neglect the effect of the non-adiabatic couplings, so the evolution becomes adiabatic and the Hamiltonian takes the form
\begin{align}\label{Eq:H_ad}
H_{\text{ad}}(t)\approx -\frac{\widetilde{\Omega}_{\text{eff}}(t)}{2}\sigma_{z}=\left[\begin{array}{cc} -\widetilde{\Omega}_{\text{eff}}(t)/2  & 0\\
0 & \widetilde{\Omega}_{\text{eff}}(t)/2 \end{array} \right].
\end{align}
As the adiabatic Hamiltonian is diagonal, there will be no population changes in the adiabatic basis,
i.e., the populations will stay constant with time and the quantum state will only accumulate a phase. We note that the effective Rabi frequency includes noisy terms
\begin{equation}
\widetilde{\Omega}_{\text{eff}}(t)=\sqrt{(\Delta(t)-\Delta_{\epsilon}(t))^2+\Omega(t)^2(1+\epsilon_{\Omega}(t))^2},
\end{equation}
which would in general cause dephasing in the adiabatic basis. However, we will show later that these can be compensated when we apply sequences of chirped adiabatic pulses.

\subsection{Mechanism of Rapid Adiabatic Passage}\label{Subsection:RAP_mechanism}

Usually, our quantum system is initially prepared (e.g., by optical pumping) with all the population in the bare state $|1\rangle$, i.e., $P_{1}(t_0)=1,~P_{2}(t_0)=0$ at the initial time $t_0$. If our goal is to transfer all the population from state $|1\rangle$ to state $|2\rangle$, i.e., $P_{1}(t_1)=0,~P_{2}(t_1)=1$ at time $t_1$ at the end of the interaction. We note that the relation between the populations of the two states and the probability amplitudes in the bare basis are given by $P_{n}(t)=|c_{n}(t)|^2$.

In order to demonstrate the mechanism for rapid adiabatic passage, we consider the composition of the probability amplitudes of the adiabatic states in terms of the ones of the bare states:
\begin{align}\label{Eq:d_amplitudes}
d_{-}(t)=c_1(t) \cos{\widetilde{\nu}(t)} - c_2(t) \sin{\widetilde{\nu}(t)}\notag\\
d_{+}(t)=c_1(t) \sin{\widetilde{\nu}(t)} + c_2(t) \cos{\widetilde{\nu}(t)}.
\end{align}
We apply an adiabatic chirped pulse from time $t_0$ to time $t_1$ where the detuning changes adiabatically from a very large negative to a very large positive value, such that
\begin{subequations}
\begin{align}
-\infty\xleftarrow{t_0\leftarrow t}&\frac{\widetilde{\Delta}(t)}{\widetilde{\Omega}(t)}\xrightarrow{t\rightarrow t_1} +\infty\\
\pi/2 = \arctan{(+\infty)}\xleftarrow{t_0\leftarrow t}&\widetilde{\nu}(t)\xrightarrow{t\rightarrow t_1} \arctan{(0)}=0\\
-c_2(t_0)\xleftarrow{t_0\leftarrow t}&d_{-}(t)\xrightarrow{t\rightarrow t_1} c_1(t_1)\\
c_1(t_0)\xleftarrow{t_0\leftarrow t}&d_{+}(t)\xrightarrow{t\rightarrow t_1} c_2(t_1).
\end{align}
\end{subequations}
It is evident that initially all the population is in state $|+\rangle$ in the dressed basis as it is aligned with state $|1\rangle$, i.e., $P_{1}(t_0)=P_{+}(t_0)=1$. As the evolution is adiabatic, the adiabatic Hamiltonian $H_{\text{ad}}(t)$ is diagonal, so there will be no transitions between the dressed states and their populations stay constant, 
so $P_{+}(t_1)=P_{+}(t_0)$. However, the mixing angle $\nu$ changes from $\pi/2$ to $0$ (see Fig. \ref{Fig1:RAP_mechanism}). As a result, the dressed state $|+\rangle$ is aligned with the state $|2\rangle$ at the final time $t_1$. Thus, $P_{2}(t_1)=P_{+}(t_1)=P_{+}(t_0)=P_{1}(t_0)=1$ and all the population is transferred adiabatically from state $|1\rangle$ to state $|2\rangle$. We note that the chirp direction is not important for the population transfer, i.e., the mixing angle can also change from $0$ to $\pi/2$ -- then the population transfer will take place via the dressed state $|-\rangle$ instead of $|+\rangle$ if the system is initially in state $|1\rangle$. It is evident that as the evolution is adiabatic, the population transfer efficiency will depend only on the initial and final values of the mixing angle and will be quite robust to amplitude and frequency fluctuations.

Figure \ref{Fig1:RAP_mechanism} shows an example for adiabatic passage with a chirped pulse for three different detuning errors $\Delta_{\epsilon}(t)$. It is evident that the population transfer efficiency is very robust to such errors even though they are greater than the peak Rabi frequency. However, the times when the flip of the quantum state takes place differ for each value of $\Delta_{\epsilon}(t)$. Specifically, the transfer process is centered at the time when the respective mixing angle is $\widetilde{\nu}(t)=\pi/4$.

One can obtain additional intuition about RAP by considering the time evolution of the energies of the adiabatic and bare states. The time evolution in the adiabatic basis leads to an avoided crossing, where the adiabatic eigenenergies $\widetilde{\epsilon}_{\pm}(t)=\pm\frac{1}{2}\sqrt{\widetilde{\Omega}(t)^2+\widetilde{\Delta}(t)^2}$ approach each other with the minimum separation at the point when $\widetilde{\Delta}(\widetilde{t})=0$ but cannot cross due to the interaction ($\widetilde{\Omega}(\widetilde{t})\ne 0$, see Fig. 2 in the main text). Meanwhile, the bare basis energies $\pm\widetilde{\Delta}(t)$ cross at a particular time $t=\widetilde{t}$, which leads to the population transfer as the mixing angle $\widetilde{\nu}(t)$ changes from $\pi/2$ to $0$. We note that adiabatic evolution is not a sufficient condition for population transfer. For example, if no crossing of the bare energies occur, the mixing angle will start at $\pi/2$ and make a return to $\pi/2$ at the end of the interaction, so we will observe a complete population return instead of complete population transfer \cite{Vitanov01ARPC}.

\subsection{Propagator of a RAP pulse}\label{Subsection:RAP_propagator}

When our goal is not simply to flip the population of the bare states, it proves useful to derive explicitly the propagator or a chirped adiabatic pulse. We consider the evolution of the system from a starting time $t_0$ to a later time $t$. In the approximation of perfect adiabaticity it is described by the propagator in the adiabatic basis
\begin{align}\label{Eq:U_ad}
U_{\text{ad}}(t,t_0)&=\exp\left[-i \left(\int_{t_0}^{t}\widetilde{\Omega}_{\text{eff}}(t^{\prime} d t^{\prime}\right)\sigma_{z}\right]\\
&=\sigma_0\cos{\left(\widetilde{\Phi}/2\right)}+i\sigma_{z}\sin{\left(\widetilde{\Phi}/2\right)},\notag
\end{align}
where
$\widetilde{\Phi}=\int_{t_0}^{t}\widetilde{\Omega}_{\text{eff}}(t^{\prime}) d t^{\prime}$ is the dynamic phase. The propagator in the bare basis then takes the form
\begin{align}\label{Eq:U_bare_RAP}
U(t,t_0)&=R_{\text{ad}}(t)^{\dagger}U_{\text{ad}}(t,t_0)R_{\text{ad}}(t_0)\\
&=\cos{\left(\widetilde{\Phi}/2\right)}\left[\sigma_0\cos{\left(\widetilde{\nu}_{r}\right)}+i\sigma_{y}\sin{\left(\widetilde{\nu}_{r}\right)}\right]\notag\\
&+i\sin{\left(\widetilde{\Phi}/2\right)}\left[\sigma_{z}\cos{\left(\widetilde{\nu}_{s}\right)}-\sigma_{x}\sin{\left(\widetilde{\nu}_{s}\right)}\right],\notag
\end{align}
where $\widetilde{\nu}_{r}\equiv \widetilde{\nu}(t)-\widetilde{\nu}(t_0)$ and $\widetilde{\nu}_{s}\equiv \widetilde{\nu}(t)+\widetilde{\nu}(t_0)$.
For example, when the evolution is perfectly adiabatic the transition probability, i.e., the probability that the qubit will be transferred to state $|2\rangle$ if it was initially in state $|1\rangle$ in the bare basis, takes the form
\begin{equation}\label{Eq:p_bare_RAP}
p=\cos{(\widetilde{\Phi}/2)}^2\cos{(\widetilde{\nu}_{r})}^2+\sin{(\widetilde{\Phi}/2)}^2\sin{(\widetilde{\nu}_{s})}^2.
\end{equation}
Assuming that we apply an adiabatic chirped pulse from time $t_0$ to time $t$ where the detuning changes from a very large negative ($\widetilde{\nu}(t_0)=\pi/2$) to a very large positive value ($\widetilde{\nu}(t)=0$), the propagator in the bare basis becomes
\begin{align}\label{Eq:U_bare}
U_{\text{RAP}}&=-i\left(\sigma_{y}\cos{(\widetilde{\Phi}/2)}+\sigma_{x}\sin{\left(\widetilde{\Phi}/2\right)}\right)\notag\\
&=\left[\begin{array}{cc} 0  & -\e^{i\widetilde{\Phi}/2}\\
\e^{-i\widetilde{\Phi}/2} & 0 \end{array} \right]
\end{align}
and the transition probability is $p=1$. In comparison, the propagator of a perfect resonant $\pi$ pulse around the $x$ axis of the Bloch sphere is $U_{\pi}=-i\sigma_{x}$. It is evident that the propagator of the perfect RAP pulse performs perfect population inversion of the bare states like a perfect $\pi$ resonant pulse but adds an additional phase rotation by the generally unknown (varying) dynamic phase $\widetilde{\Phi}+\pi$ in the xy plane of the Bloch sphere of the qubit.

\subsection{Conditions for Rapid Adiabatic Passage}\label{Section:RAP_conditions}

There are two main conditions for RAP and we will discuss them separately. First, the evolution should be adiabatic, so no transitions take place in the adiabatic basis. Second,
the mixing angle $\widetilde{\nu}(t)$ should change from $\pi/2$ to $0$ (or vice versa).

\subsubsection{Adiabatic condition}\label{Subsection:Adiab_condition}

The first requirement is that the non-adiabatic coupling is much smaller than the energy separation between the adiabatic states, so no transitions occur
\begin{equation}
\frac{|\widetilde{\nu}^{\prime}(t)|}{\epsilon_{+}(t)-\epsilon_{-}(t)}\ll 1,
\end{equation}
which can be simplified to \cite{Vitanov01ARPC}
\begin{equation}\label{Eq:Adiab_cond_general}
\frac{\left|\dot{\widetilde{\Omega}}(t)\widetilde{\Delta}(t)-\widetilde{\Omega}(t)\dot{\widetilde{\Delta}}(t)\right|}{2\left(\widetilde{\Delta}(t)^2+\widetilde{\Omega}(t)^2\right)^{3/2}}\ll 1.
\end{equation}
The exact formula for this condition depends on the specific time-dependence of $\widetilde{\Omega}(t)$ and $\widetilde{\Delta}(t)$. Usually, adiabaticity is worst at the moment of level crossing of the bare energies, i.e., when $\widetilde{\Delta}(t_{\text{c}})=0$, so it is determined by the element $\widetilde{\Omega}(t)\dot{\widetilde{\Delta}}(t)$ in the numerator in Eq. \eqref{Eq:Adiab_cond_general}.
We note that when the chirp range is small (but non-zero), e.g., of the order of the peak Rabi frequency, the element $\dot{\widetilde{\Omega}}(t)\widetilde{\Delta}(t)$ can become significant for certain pulse shapes. However, we are usually be interested in the case of smooth pulses when peak Rabi frequency is too weak to cover the inhomogeneous broadening, which requires a large chirp range. Then, $\widetilde{\Omega}(t)\dot{\widetilde{\Delta}}(t)$ is dominant and
the condition simplifies to
\begin{equation}
\frac{|\dot{\widetilde{\Delta}}(t_{\text{c}})|}{2\widetilde{\Omega}(t_{\text{c}})^2}\ll 1,
\end{equation}
or equivalently to the so called lower boundary adiabatic condition
\begin{equation}
\frac{\widetilde{\Omega}(t_{\text{c}})^2}{|\dot{\widetilde{\Delta}}(t_{\text{c}})|}\gg 1,
\end{equation}
where $\dot{\widetilde{\Delta}}(t_{\text{c}})$ is the chirp rate at the time $t_{\text{c}}$ of the crossing of the bare states energies ($\Delta(t_{\text{c}})=0$).
Figure \ref{FigX:RAP_adiab_simulation} includes an example for the lower boundary condition, which shows 
that even moderate levels of the order of 3.3 are enough to reach transition probabilities of the order of 0.9.
As the chirp rate is usually bounded by the chirp range, given a fixed pulse duration, this requirement imposes a condition for a maximum chirp range.

\subsubsection{Condition for mixing angle evolution}\label{Subsection:Mixing_angle}

The second condition for population transfer requires that the mixing angle $\widetilde{\nu}(t)$ changes from $\pi/2$ to $0$ (or vice versa), which in turn imposes a condition on a minimum chirp range.
In case of perfect adiabaticity, it can be shown that the transition probability in Eq. \eqref{Eq:p_bare_RAP} can also be presented as \cite{Vitanov01ARPC}
\begin{equation}
p=\frac{1}{2}-\frac{\widetilde{\Delta}(t_1)\widetilde{\Delta}(t_0)}{2\widetilde{\Omega}_{\text{eff}}(t_1)\widetilde{\Omega}_{\text{eff}}(t_0)}-\frac{\widetilde{\Omega}(t_1)\widetilde{\Omega}(t_0)}{2\widetilde{\Omega}_{\text{eff}}(t_1)\widetilde{\Omega}_{\text{eff}}(t_0)}\cos{\widetilde{\Phi}},
\end{equation}
This expression can be simplified further if we assume that the magnitude of the Rabi frequency at the beginning and the end of the interaction is much smaller than the detuning and thus than the effective Rabi frequency, i.e.,  $\widetilde{\Omega}(t_{k})\ll\widetilde{\Omega}_{\text{eff}}(t_{k}),~k=0,1$. Then, we can neglect the fast-oscillating last term and obtain
\begin{equation}
p\approx\frac{1}{2}-\frac{\widetilde{\Delta}(t_1)\widetilde{\Delta}(t_0)}{2\widetilde{\Omega}_{\text{eff}}(t_1)\widetilde{\Omega}_{\text{eff}}(t_0)}\approx\frac{1}{2}\left(1+\frac{\widetilde{\Delta}(t_1)^2}{\widetilde{\Omega}_{\text{eff}}(t_1)^2}\right),
\end{equation}
where we assumed in the last equality that the Rabi frequency is a symmetric function with respect to the center of the pulse, i.e.
$\widetilde{\Omega}(t_1)\approx\widetilde{\Omega}(t_0)$,
and the detuning is an anti-symmetric function, so $\widetilde{\Delta}(t_1)\approx-\widetilde{\Delta}(t_0)$. This is a feasible assumption if the magnitude of target detuning $|\Delta(t)|\gg |\Delta_{\epsilon}|$ at the beginning and the end of the interaction. Thus, we obtain
\begin{equation}
p\approx 1-\frac{1}{2}\frac{\widetilde{\Omega}(t_1)^2}{\widetilde{\Omega}_{\text{eff}}(t_1)^2}=
1-\frac{2}{4+(R/\widetilde{\Omega}(t_1))^2},
\end{equation}
where $R\approx\widetilde{\Delta}(t_1)-\widetilde{\Delta}(t_0)\approx 2|\widetilde{\Delta}(t_1)|$ is the magnitude of the target chirp range. It is evident that perfect population transfer requres that the ratio $R/\widetilde{\Omega}(t_1)\gg 1$. If we require the error in the population transfer efficiency $\epsilon\equiv 1-p\le \epsilon_{\text{max}}$, the condition becomes
\begin{equation}\label{Eq:chirp_rate_condition}
\frac{R}{\widetilde{\Omega}(t_1)}\ge \sqrt{\frac{2}{\epsilon_{\text{max}}}-4},
\end{equation}
where $\epsilon_{\text{max}}$ is the maximum error in the transfer efficiency, which we assumed to be $\epsilon\le 1/2$ by requiring that the initial and final detunings have opposite signs. For example, an error in the population transfer efficiency of $\epsilon\le 0.1$ requires $R/\widetilde{\Omega}(t_1)\ge 4$, while $\epsilon\le0.01$ implies $R/\widetilde{\Omega}(t_1)\ge 14$.

We note that the condition $\widetilde{\Delta}(t_1)\approx-\widetilde{\Delta}(t_0)$ might not be satisfied in systems with large inhomogeneous broadening and the chirp range requirement needs to be modified to cover the shift in the initial and final detuning. If $\widetilde{\Delta}(t_0)=-(R/2)+\Delta_{\epsilon}$ and $\widetilde{\Delta}(t_1)=(R/2)+\Delta_{\epsilon}$, we obtain
\begin{align}
p&\approx\frac{1}{2}-\frac{\widetilde{\Delta}(t_1)\widetilde{\Delta}(t_0)}{2\widetilde{\Omega}_{\text{eff}}(t_1)\widetilde{\Omega}_{\text{eff}}(t_0)}\notag\\
&=\frac{1}{2}\left(1+\frac{1-x^2}{\sqrt{\left((1-x)^2+y^2\right)\left((1+x)^2+y^2\right)}}\right),
\end{align}
where $x\equiv 2\Delta_{\epsilon}/R$ and $y\equiv 2\widetilde{\Omega}(t_1)/R$ and we assumed that $\widetilde{\Omega}(t_0)=\widetilde{\Omega}(t_1)$. We require $\Delta_{\epsilon}<R/2$ in order for the detuning $\Delta_{\epsilon}$ to lie within the chirp range, which implies $x<1$. Usually, the Rabi frequency is much smaller than the detuning at the beginning and the end or the pulse, so in the approximation $y\to 0$ the error in the transition probability becomes
\begin{equation}
\epsilon\approx \frac{1+x^2}{2(1-x^2)^2}y^2.
\end{equation}
This implies that for $\epsilon \le \epsilon_{\text{max}}$, we require
\begin{equation}\label{Eq:chirp_rate_condition_with_detuning}
\frac{R}{\widetilde{\Omega}(t_1)}\ge \frac{1}{1-x^2}\sqrt{\frac{2(1+x^2)}{\epsilon_{\text{max}}}}.
\end{equation}
The formula converges to the one in Eq. \eqref{Eq:chirp_rate_condition} when $x=0$ and in the limit $\epsilon_{\text{max}}\to 0$. For example, when $\Delta_{\epsilon}/R=0.25$, i.e., $x=0.5$, the error $\epsilon\le0.01$ implies $R/\widetilde{\Omega}(t_1)\ge 21.1$, which is higher than the value of $14$ in the noiseless case. Thus, in the presence of detuning errors, we require a larger ratio of $R/\widetilde{\Omega}(t_1)$ to reach the same transfer efficiency.

In summary, the adiabatic condition requires a small chirp rate (and thus a small chirp range) while the condition that the mixing angle $\widetilde{\nu}(t)$ changes from $\pi/2$ to $0$ (or vice versa) requires a large chirp range. Next, we show an example for RAP conditions for the pulse shape of the Allen-Eberly model \cite{HioePRA1984,Allen-Eberly1987}, which we use in the manuscript.

\subsection{Example: Allen-Eberly model}\label{Subsection:AE}

We describe the conditions for RAP for the Allen-Eberly (AE) model \cite{Allen-Eberly1987,HioePRA1984}, which is characterised by a Rabi frequency and a detuning with the following shapes (see Fig. 2 (top) in the main text)
\begin{subequations}\label{Eq:AE_model_appendix}
\begin{align}
\Omega(t)&=\Omega_0~\sech{(t/T)}\\
\Delta(t)&=\Delta_0 \tanh{(t/T)},~t\in \left[-T_{\text{pulse}}/2,T_{\text{pulse}}/2\right],
\end{align}
\end{subequations}
where $T$ is a characteristic time of the RAP pulse and $T_{\text{pulse}}$ is the RAP pulse duration and we dropped the noisy terms for simplicity of presentation and assumed that the pulse is centered at time $t_{\text{c}}=0$.

The lower boundary adiabatic condition for this model simplifies to
\begin{equation}\label{Eq:Adiab_AE}
\frac{\Omega(t_{\text{c}})^2}{\dot{\Delta}(t_{\text{c}})}=\frac{\Omega_0^2 T}{\Delta_0}\gg 1,
\end{equation}
where $t_{\text{c}}$ is the moment of level crossing ($\Delta(t_{\text{c}})=0$). Equivalently, this criterion can be given in terms of the target chirp range $R=2\Delta_0$:
\begin{equation}
\frac{R}{\Omega_0}\ll \frac{\Omega_0 T}{2}.
\end{equation}
We note that when the RAP pulse duration $T_{\text{pulse}}\to\infty$, the pulse area $A=\int_{-T_{\text{pulse}}/2}^{T_{\text{pulse}}/2}\Omega(t) d t=\pi \Omega_0 T$, so the ratio between the maximum chirp range and the peak Rabi frequency is simplified to $R/\Omega_0\ll 2A/\pi$. 
Another important advantage of the AE model in comparison to the standard Landau-Zener-Stückelberg-Majorana model with a constant drive and a linear chirp \cite{LandauZener1932}  is that the pulse area (and thus the energy input into the system) is limited, no matter how long is the pulse duration $T_{\text{pulse}}$.

\red
\begin{figure}[t!]
\includegraphics[width=\columnwidth]{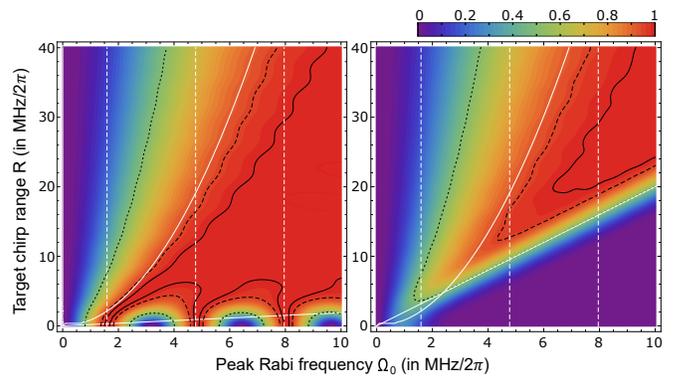}
\caption{(color online)
Numerical simulation of single RAP pulse transition probability vs. peak Rabi frequency $\Omega_0$ and target chirp range $R$. We assume a two-state quantum system interacting with a driving field with a Rabi frequency $\widetilde{\Omega}(t)=\Omega_0 \sech{\left(\frac{t-t_{c}}{T}\right)}$ and detuning $\widetilde{\Delta}(t)=\Delta_{\epsilon}+(R/2)\tanh{\left(\frac{t-t_{c}}{T}\right)}$ with pulse duration $T_{\text{pulse}}=1~\mu$s and characteristic time $T=0.1~\mu$s for (left) zero static detuning $\Delta_{\epsilon}=0$ and (right) $\Delta_{\epsilon}=\Omega_0$. The black dotted/dashed/solid lines correspond to $p$ of $0.5/~0.95/~0.99$, respectively. The vertical, white, dashed lines correspond to pulse areas of $\pi$, $3\pi$, $5\pi$. The top solid white line corresponds to the adiabaticity parameter $\widetilde{\Omega}(t_{\text{c}})^2/|\dot{\widetilde{\Delta}}(t_{\text{c}})|=3.33$ (see Eq. \eqref{Eq:Adiab_AE}). The bottom solid line corresponds to (a) the chirp range condition $R\sinh{(T_{\text{pulse}}/2T)}= \Omega_0\sqrt{\frac{2}{\epsilon_{\text{max}}}-4}$ for $\epsilon_{\text{max}}=0.01$ (see Eq. \eqref{Eq:chirp_rate_condition}), and (b) $R=2\Delta_{\epsilon}$. As expected from theory, the good performance of RAP pulses takes place in the region between the white solid lines and can be achieved even for moderate pulse areas.
}
\label{FigX:RAP_adiab_simulation}
\end{figure}
\black

Next, we consider the condition for mixing angle evolution for this model. First, we note that the actual chirp range is given by $\Delta(T_{\text{pulse}}/2)-\Delta(-T_{\text{pulse}}/2)=R\tanh{(T_{\text{pulse}}/2T)}$ and approaches the target chirp range only when $(T_{\text{pulse}}/T\to\infty)$.
We note that this is not very restrictive
but one needs an interaction time of several times T to ensure that the actual chirp range is similar to the maximum one and there is negligible truncation of the Rabi frequency function. For example, the truncation is quite small for $T_{\text{pulse}}/T= 10$ when $\Omega(T_{\text{pulse}}/2)\approx 0.013 \Omega_0$ and the actual chirp range is $\approx 0.9999 R$.
The condition for mixing angle evolution is then given in Eq. \eqref{Eq:chirp_rate_condition} and takes the form
\begin{equation}
\frac{R\sinh{(T_{\text{pulse}}/2T)}}{\Omega_0}\ge \sqrt{\frac{2}{\epsilon_{\text{max}}}-4}.
\end{equation}
%
An example of the relevance of the mixing angle condition as a lower boundary of the chirp range is given in Fig. \ref{FigX:RAP_adiab_simulation}(a).
Then, both conditions can be summarized to obtain the following double inequality for the ratio between the target chirp range and the peak Rabi frequency
\begin{equation}\label{Eq:AE_conditions}
\frac{1}{\sinh{(T_{\text{pulse}}/2T)}}\sqrt{\frac{2}{\epsilon_{\text{max}}}-4}\le \frac{R}{\Omega_0}\ll \frac{2}{\pi}A.
\end{equation}
Thus, RAP requires a minimum ratio of the chirp range and the peak Rabi frequency $(R/\Omega_0)$ to ensure sufficient change in the mixing angle to ensure a transition probability error no greater than $\epsilon_{\text{max}}$ (left inequality). Additionally, the ratio $(R/\Omega_0)$ should be much smaller than the pulse area $A$ to ensure adiabaticity (right inequality).
The relevance of both conditions is demonstrated in Fig. \ref{FigX:RAP_adiab_simulation}(a), where the region of high transition probability lies between the white solid lines that describe them.
We note that this model requires a smaller ratio between the maximum chirp range and the peak Rabi frequency than the standard Landau-Zener-Stückelberg-Majorana model with a constant drive. This, in turn, leads to lower requirements for pulse area (and energy input) although the interaction time $T$ can be kept the same.

Next, we note that this particular model can be solved analytically in the limit when $T_{\text{pulse}}/T\to\infty$ even without assuming adiabaticity, giving the transfer efficiency \cite{Kyoseva06PRA}
\begin{equation}\label{Eq:AE_population}
p\to 1-\sech{\left(\frac{\pi T R}{4}\right)}^2\cos{\left( \frac{\pi T \sqrt{|R^2-4\Omega_0^2|}}{4}\right)}^2.
\end{equation}

Finally, we discuss briefly the case when an additional detuning error is present, e.g., due to inhomogeneous broadening. Then, $\widetilde{\Delta}(t)=\Delta_{\epsilon}+\Delta(t)$ and the dynamics are more complex. Then, the time of the level crossing is shifted by $T \arctan{(-2\Delta_{\epsilon}/R)}$ in comparison to the noiseless case due to the detuning error (see Fig. \ref{Fig1:RAP_mechanism}). The lower boundary adiabaticity condition and, thus, the right inequality in Eq. \eqref{Eq:AE_conditions}, is not affected by this shift for the AE model since the ratio $\Omega(t)^2/\dot{\Delta}(t)=2\Omega_0^2 T/R$ is independent of $t$, i.e., the moment of the level crossing
(see Fig. \ref{FigX:RAP_adiab_simulation}).
We note that this would usually not the case for other pulse shapes.

However, we need to modify the requred chirp range in accordance to Eq. \eqref{Eq:chirp_rate_condition_with_detuning} and obtain
\begin{equation}\label{Eq:AE_conditions_with_detuning}
\frac{1}{\sinh{(T_{\text{pulse}}/2T)}}\frac{1}{1-\widetilde{x}^2}\sqrt{\frac{2(1+\widetilde{x}^2)}{\epsilon_{\text{max}}}}\le \frac{R}{\Omega_0}\ll \frac{2}{\pi}A,
\end{equation}
where $\widetilde{x}=2\Delta_{\epsilon}/(R\tanh{(T_{\text{pulse}}/2T)})$. In other words, one has to apply a slightly longer pulse or increase the chirp range to reach the same transfer efficiency as in the noiseless case.

We note that the particular model with additional static detuning, such that $\Delta(t)=\Delta_{\epsilon}+\Delta_0 \tanh{(t/T)}$ can be solved analytically in the limit when $T_{\text{pulse}}/T\to\infty$ even without assuming adiabaticity. It is then termed Demkov-Kunike model and the transfer efficiency is \cite{Kyoseva06PRA}
\begin{align}\label{Eq:DK_population}
p &\to 1-|\widetilde{\epsilon}|^2,~\text{where}\notag\\
\widetilde{\epsilon} &=\frac{\Gamma\left(\frac{1}{2}+i(\delta+\chi)\right)\Gamma\left(\frac{1}{2}+i(\delta-\chi)\right)}
{\Gamma\left(\frac{1}{2}+\sqrt{\alpha^2-\chi^2}+i\delta\right)\Gamma\left(\frac{1}{2}-\sqrt{\alpha^2-\chi^2}+i\delta\right)},\notag\\
\alpha &=\Omega_0 T/2,~\delta=\Delta_{\epsilon}T/2,~\chi=\Delta_0 T/2.
\end{align}

\subsection{RAP transition time}\label{Subsection:RAP_transition_time}

\begin{figure}[t!]
\includegraphics[width=\columnwidth]{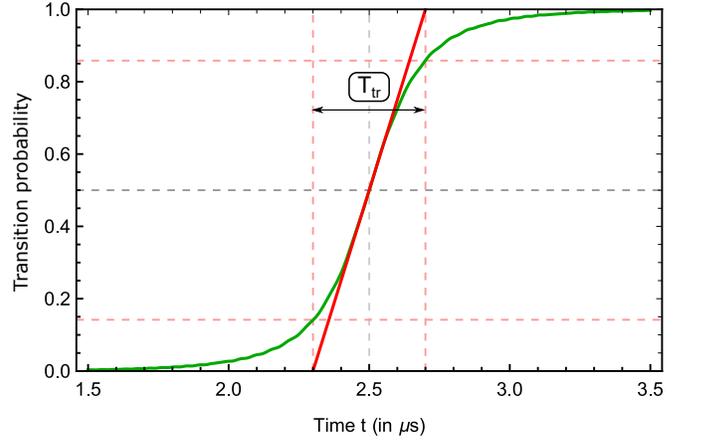}
\caption{(color online)
Example of transition time $T_{\text{tr}}$ in RAP. We perform a numerical simulation of a two-state quantum system interacting with a driving field with a Rabi frequency $\widetilde{\Omega}(t)=\Omega_0 \sech{\left(\frac{t-t_{c}}{T}\right)}$, where the peak Rabi frequency $\Omega_0=2\pi~10$ MHz and $T=0.5~\mu$s, detuning $\widetilde{\Delta}(t)=(R/2)\tanh{\left(\frac{t-t_{c}}{T}\right)}$, where $R=2\pi~50$ MHz is the target chirp range, the initial and end times are $t=0$ and $t=T_{\text{pulse}}=5~\mu$s. We show the time evolution of (green) the transition probability $p(t)$ in the bare states basis, (red) the function
$|\widetilde{\nu}^{\prime}(t_{c})|(t-t_{c})$, where $t_{c}=2.5~\mu$s is the center of the pulse. The transition time is $T_{\text{tr}}=4\Omega_0 T/R=0.4~\mu$s. As expected from theory, the transition probability at $t=t_{c}+T_{\text{tr}}/2$ is $p=(2+\sqrt{2})/4\approx 0.854$.}
\label{Fig1:RAP_transition_time}
\end{figure}

We discuss now the transition time in RAP, i.e., this is the characteristic time, which describes the duration of the population transfer from state $|1\rangle$ to state $|2\rangle$ in RAP.
We use a definition of transition time, which was proposed previously by Boradjiev et. al. \cite{Boradjiev2010PRA} in the context of stimulated Raman adiabatic passage. The transition time is defined as
\begin{equation}\label{Eq:Transition_time_def}
T_{\text{tr}}=\frac{1}{|\partial_{t}p(\nu(t_{c}))|}
=\frac{1}{|\widetilde{\nu}^{\prime}(t_{c})|}=\left|\frac{2\widetilde{\Omega}(t_{c})}{\widetilde{\Delta}^{\prime}(t_{c})}\right|,
\end{equation}
where $t_{c}$ is the time when the detuning $\widetilde{\Delta}(t)$ crosses resonance and the mixing angle becomes $\widetilde{\nu}(t_{c})=\pi/4$. The transition time is inversely proportional to the non-adiabatic coupling at the moment of level-crossing in the bare basis 
and depends on the specific model for RAP. In case of the Allen-Eberly model \cite{Allen-Eberly1987} in Eq. \eqref{Eq:AE_model_appendix}, the transition time takes the form
\begin{equation}\label{Eq:Transition_time_special}
T_{\text{tr}}=\frac{4\Omega_0 T}{R},
\end{equation}
where $\Omega_0$ is the peak Rabi frequency, reached at time $t_{c}$ of the level-crossing in the bare basis, $R$ is the maximum chirp range, and we assumed no noise.
In the presence of noise, e.g., due to inhomogeneous broadening, we observe a shift in the detuning to $\widetilde{\Delta}(t)=\Delta_{\epsilon}+\Delta(t)$, which changes the moment when the $\widetilde{\Delta}(t)$ crosses resonance. Then, the Rabi frequency can be lower than its peak value and the derivative of $\widetilde{\Delta}(t)$ can also differ, which modifies the transition time. For the example model in Eq. \eqref{Eq:AE_model_appendix}, which we use in our work, the moment of resonance crossing is shifted by $T \arctan{(-2\Delta_{\epsilon}/R)}$ in comparison to the noiseless case and the modified transition time becomes
\begin{equation}\label{Eq:Transition_time_special}
T_{\text{tr}}=\frac{4\Omega_0 T}{\sqrt{R^2-4\Delta_{\epsilon}^2}}.
\end{equation}

The transition probabilities at times $t_{c}\pm m T_{\text{tr}}/2$ can be calculated exactly for this model and are given by
\begin{equation}\label{Eq:Transition_prob_Ttr}
p=\frac{1}{2}\pm\frac{1}{2}\left(1+\frac{k^2}{\sinh{(mk)}^2}\right)^{-1/2},~k\equiv 2\Omega_0/R,
\end{equation}
where $m\ge 0$. The lower bound of the transition probability is achieved for $k\to 0$, i.e., infinitely large chirp range with respect to the peak Rabi frequency, and takes the form
\begin{equation}\label{Eq:Transition_prob_Ttr_min}
p_{\text{min}}=\frac{1}{2}\left(1\pm\frac{m}{\sqrt{1+m^2}}\right),
\end{equation}
Thus, the lower bound of the transition probability for $m=1$, i.e., at time $t_{c}+T_{\text{tr}}/2$, is $p_{\text{min}}=(2+\sqrt{2})/4\approx 0.854$, while for $m=2$: $p_{\text{min}}=1/2+1/\sqrt{5}\approx 0.947$.

\section{Robust sequences of RAP pulses}\label{Section:Robust_RAP_sequences}

\subsection{Dynamic phase compensation}\label{Subsection:Dynamic_phase}

We consider sequences of RAP pulses in this section.
First, we note that the dynamic phase $\widetilde{\Phi}$ of a RAP pulse can be compensated completely when we apply two RAP pulses, as long as it is the same during the first and the second pulses and they perform perfect population inversion. Then, the propagator in the bare basis is $U_{\text{RAP}}U_{\text{RAP}}=-\sigma_0$, where $\sigma_0$ is the identity matrix, which is independent from $\widetilde{\Phi}$.

It proves useful to consider the compensation mechanism by analyzing the evolution in the adiabatic basis when we apply two RAP pulses. During the first RAP pulse from time $t_0$ to time $t_1$ the Hamiltonian in the adiabatic basis is given by Eq. \eqref{Eq:H_ad_no_approx_appendix}. We then assume for simplicity that there is no pulse separation between the RAP pulses. Then, at the start of the second RAP pulse, we need to apply a very fast, (approximately) instantaneous change in the sign of the target detuning from $\Delta(t)\to -\Delta(t)$. This leads to a sudden change in the mixing angle from $0$ to $\pi/2$, i.e., $\Delta\widetilde{\nu}=\pi/2$. The Hamiltonian in the adiabatic basis during this change is dominated by the non-adiabatic coupling and is given by $H_{\text{ad}}(t)\approx\widetilde{\nu}^{\prime}(t)\sigma_{y}$,
where $\widetilde{\nu}^{\prime}(t)$ has an (approximately) delta function behavior and its integral is the change in the mixing angle $\Delta\widetilde{\nu}=\pi/2$. Thus, the evolution in the adiabatic basis in the infinitesimal time between the two RAP pulses is given by the propagator
$
\exp{(-i\Delta\widetilde{\nu}\sigma_{y}})=-i\sigma_{y}.
$
Thus, the adiabatic states are interchanged. It is evident that the sudden change in the mixing angle plays the role of a $\pi$ pulse around the $y$ axis in the adiabatic basis. As a result, the phase evolution during the second RAP pulse compensates the one during the first RAP pulse, as long as the accumulated dynamic phase is the same during both RAP pulses.


We can incorporate the transitions due to the sudden changes of the mixing angle in the basis itself. Thus, we can define a new basis, which we term \emph{``adiabatic, toggling''} basis. The transformation matrix from the adiabatic to the ``adiabatic, toggling'' basis for times during the $k$-th RAP pulse is given by
\begin{align}\label{Eq:R_adiabatic_tog}
R_{\text{ad,tog}}(t)&=\exp{\left[i(k-1)\Delta\widetilde{\nu}\sigma_{y}\right]},~t\in(t_{k-1},t_{k})\notag\\
H_{\text{ad,tog}}(t)&=R_{\text{ad,tog}}(t)H_{\text{ad}}(t)R_{\text{ad,tog}}(t)^{\dagger}
\end{align}
where $t_{k-1}$ is the beginning and $t_{k}$ is the end of the $k$-th RAP pulse. The Hamiltonian in the ``adiabatic, toggling'' basis in the adiabatic approximation then takes the form
\begin{align}\label{Eq:H_adiabatic_tog}
H_{\text{ad,tog}}(t)&= f(t)\widetilde{\Omega}_{\text{eff}}(t)\sigma_{z}/2,
\end{align}
where $f(t)=-1$ during the odd-numbered RAP pulses and $f(t)=1$ during the even-numbered ones, and we assumed that $\Delta\widetilde{\nu}=\pi/2$ between two RAP pulses. Thus, the accumulated dynamic phase, including the effect of the frequency and amplitude noise, is compensated during every second RAP pulse as long as it is the same as in the previous pulse, i.e., the correlation time of the noise is long in comparison to the duration of two RAP pulses. 

\subsection{Phased sequences of RAP pulses}\label{Subsection:Phased_RAP}

Perfect RAP pulses are difficult to achieve in real experimental realizations because the adiabaticity condition is hard to fulfill and/or the mixing angle might take a very long time to change from $\widetilde{\nu}(t_0)=\pi/2$ to $\widetilde{\nu}(t_1)=0$ during a single RAP pulse. In order to compensate these errors we will use the relative phases of the RAP pulses $\phi_{k}$ as additional control parameters and apply robust sequences of pulses. For example, we can choose the phases of the individual RAP pulses to correspond to the popular XY, KDD, or UR sequences \cite{RDD_review12Suter,Torosov11PRL,Genov2014PRL,GenovPRL2017}. These are based on composite pulses, which have been shown to improve the efficiency of population transfer and rephasing with imperfect RAP pulses \cite{Torosov11PRL,Schraft13PRA,GenovPRL2017,Genov2018PRA}.

The propagator of a pulse (not necessarily RAP) in the bare basis can be parameterized by \cite{Genov2014PRL,GenovPRL2017,Genov2018PRA}
\begin{equation} \label{Eq:U_bare}
\mathbf{U} = \left[\begin{array}{cc} \sqrt{\epsilon} \e^{i\alpha}  & \sqrt{1-\epsilon} \e^{-i\beta} \\  -\sqrt{1-\epsilon} \e^{i\beta} & \sqrt{\epsilon} \e^{-i\alpha}  \end{array} \right],
\end{equation}
where $p\equiv 1-\epsilon$ is the transition probability, i.e., the probability that the qubit will be transferred to state $|2\rangle$ if it was initially in state $|1\rangle$, $\epsilon\in[0,1]$ is the unknown error in the transition probability, $\alpha$ and $\beta$ are unknown phases. For example, when the evolution is perfectly adiabatic the transition probability is given by Eq. \eqref{Eq:p_bare_RAP}.
In case of a perfect RAP pulse, 
the transition probability becomes $p=1$ and $\epsilon=0$. However, this is often not the case, e.g., due to imperfect adiabaticity or insufficient change in the mixing angle during a RAP pulse.
For example, imperfect adiabaticity due to, e.g., fast change in the mixing angle or insufficient pulse area, can make $\epsilon\ne 0$. Such errors can be compensated by applying phased sequences of pulses, where the phases of the subsequent pulses are chosen to cancel the errors of the individual pulses up to a certain order \cite{Genov2014PRL,GenovPRL2017,Genov2018PRA}.
5

If the pulses are time separated, the propagator of the whole cycle $\left[\right.$free evolution for time $\tau/2-\text{pulse}-$ free evolution for time $\tau/2\left.\right]$ changes by taking $\alpha\to\widetilde{\alpha}=\alpha+\Delta_{\epsilon}\tau$, where we assumed that the detuning variation $\Delta_{\epsilon}$ is constant during one $\left[\tau/2-\text{pulse}-\tau/2\right]$ period. Additionally, a shift in the phase $\phi_{k}$ at the beginning of a pulse (see Eq. \eqref{Eq:H_general}) causes $\beta\to\beta+\phi_{k}$ \cite{Genov2014PRL,GenovPRL2017}. Thus, the propagator of the $k$-th pulse in the bare basis takes the form
\begin{align}\label{Eq:U_bare_phased}
U(\phi_k)=\left[\begin{array}{cc} \sqrt{\epsilon}\e^{i\widetilde{\alpha}}  & \sqrt{1-\epsilon}\e^{-i(\beta+\phi_{k})}\\
-\sqrt{1-\epsilon}\e^{i(\beta+\phi_{k})} &\sqrt{\epsilon}\e^{-i\widetilde{\alpha}} \end{array} \right].
\end{align}
Assuming coherent evolution during a sequence of $n$ pulses with different initial phases $\phi_{k}$, the propagator of the composite sequence then becomes
\begin{equation}\label{Eq:U_phased_sequence}
U^{(n)}=U(\phi_{n})\dots U(\phi_{1}),
\end{equation}
and the phases $\phi_{k}$ of the individual pulses can be used as control parameters to achieve a robust performance.
We can evaluate the latter by considering the fidelity \cite{RDD_review12Suter,GenovPRL2017}
\begin{equation}\label{Eq:Fid_phased_sequence}
F=\frac{1}{2}\text{Tr}\left[\left(U_0^{(n)}\right)^{\dagger}U^{(n)}\right],
\end{equation}
where $U_0^{(n)}$ is the propagator of the respective pulse sequence when $\epsilon=0$, i.e., when the pulse performs a perfect population inversion. For example, the fidelity of a single pulse is given by $F=\sqrt{1-\epsilon}$. We note that this measure of fidelity does not take into account variation in the phase $\beta$, which is important when we apply an odd number of pulses. However, the latter is fully compensated when we apply an even number of pulses with perfect transition probability. Thus, we use the fidelity measure in Eq. \eqref{Eq:Fid_phased_sequence} as it usually provides a simple and sufficient measure of performance when we apply an even number of pulses.
We can obtain the fidelity of a
sequence of eight pulses with zero phases, i.e., $\phi_{k}=0$, which is given by
\begin{equation}\label{Eq:Fid_zero8}
F_{(\phi_{k}=0)}=1-32 \cos{(\widetilde{\alpha})}^4\epsilon-O(\epsilon^2).
\end{equation}
Additionally, the fidelity of the widely used XY8 sequence \cite{RDD_review12Suter} with phases $\left(0,1,0,1,1,0,1,0\right)\pi/2$ is
\begin{equation}\label{Eq:Fid_zero8}
F_{\text{XY8}}=1-4\left[\cos{(\widetilde{\alpha})}+\cos{(3\widetilde{\alpha})}\right]^2\epsilon^3-O(\epsilon^4).
\end{equation}
Usually the transition probability error is quite small, i.e., $\epsilon \to 0$, so the error in the fidelity ($1-F$) of the XY8 sequence ($\sim \epsilon^3$) will be much smaller than the one of the sequence with constant zero phases ($\sim \epsilon$). Similarly, one can show that we can obtain a robust performance and even better fidelity with other sequences of phased pulses, e.g., by using the KDD or UR sequences \cite{RDD_review12Suter,Genov2014PRL,GenovPRL2017}

We note that we made no assumption of the pulse shape and detuning time dependence during this analysis, except for the RWA to obtain the Hamiltonian in Eq. \eqref{Eq:H_general_RWA}, coherent evolution, and the assumption that effect of the pulse and free evolution before and after the pulse on the qubit is the same during each pulse (except for the effect of the phase $\phi_{k}$). Thus, the analysis is applicable for sequences of RAP pulses \cite{Genov2018PRA}. We note that when the detuning $\widetilde{\Delta}(t)$ is an antisymmetric function of time with respect to the center of a RAP pulse (e.g., when $\Delta_{\epsilon}=0$), the phase $\widetilde{\alpha}=0$, which allows for additional simplification, as used in \cite{Torosov11PRL}.

\section{Detailed theory of RAP sensing}\label{Section:RAP_sensing}

In this section we show how we can apply RAP for sensing. Our goal is to sense the amplitude of an oscillating (AC) field. 
We consider the Hamiltonian
\begin{align}
\widetilde{H}_{\text{s}}(t)&=\frac{\widetilde{\omega}_0(t)}{2}\sigma_{z}+\widetilde{\Omega}(t)\sigma_{x}\cos{\left[\omega_0 t +\delta(t)+\phi\right]}\notag\\
&+g\sigma_{z}\cos{(\omega_{\text{s}}t+\xi)},
\end{align}
where $g$ is the amplitude of the oscillating sensed field, $\omega_{\text{s}}$ is its angular frequency and $\xi$ is its initial phase.

We move to the rotating frame with respect to the carrier frequency $\omega(t)\sigma_{z}/2$, apply the rotating-wave approximation ($|\Omega(t)|\ll\omega$) and obtain the Hamiltonian
\begin{align}
H_{\text{s}}(t)&=-\frac{\widetilde{\Delta}(t)}{2}\sigma_{z}+\frac{\widetilde{\Omega}(t)}{2}\sigma_{x}+g\sigma_{z}\cos{(\omega_{\text{s}}t+\xi)},
\end{align}
where we took $\phi=0$ without loss of generality.
We now move to the adiabatic basis, as defined in sec. \ref{Section:RAP_Theory}. The Hamiltonian takes the form
\begin{align}\label{Eq:H_ad_sense}
H_{\text{ad,s}}(t)&= -\frac{\widetilde{\Omega}_{\text{eff}}(t)}{2}\sigma_{z}
+g\cos{(\omega_{\text{s}}t+\xi)}\\
&\times\left[\cos{(2\widetilde{\nu}(t))}\sigma_{z}+\sin{(2\widetilde{\nu}(t))}\sigma_{x}\right],\notag
\end{align}
where we applied the adiabatic approximation, assuming
$
|\widetilde{\nu}^{\prime}(t)|\ll\widetilde{\Omega}_{\text{eff}}(t).
$
It proves useful to incorporate any instantaneous changes to the mixing angle by moving to the ``adiabatic, toggling'' basis, as defined in Eq. \eqref{Eq:R_adiabatic_tog}, where the Hamiltonian becomes
\begin{align}\label{Eq:H_adiabatic_tog_sensing}
H_{\text{ad,tog,s}}(t)&=f(t)\frac{\widetilde{\Omega}_{\text{eff}}(t)}{2}\sigma_{z}
-f(t)g \cos{(\omega_{\text{s}}t+\xi)}\notag\\
&\times\left[\cos{(2\widetilde{\nu}(t))}\sigma_{z}+\sin{(2\widetilde{\nu}(t))}\sigma_{x}\right].
\end{align}
If no sudden changes in the mixing angle occur, the ``adiabatic, toggling'' basis is the same as the standard adiabatic basis and $f(t)=-1$. If we apply sequences of RAP pulses where $\Delta\nu=\pi/2$ between the pulses, then $f(t)=-1$ during the odd-numbered RAP pulses and $f(t)=1$ during the even-numbered ones.
Finally, we move to the interaction basis with respect to $f(t)\widetilde{\Omega}_{\text{eff}}(t)/2$ and obtain the Hamiltonian
\begin{align}\label{Eq:H_ad_sense_interaction_tog}
H_{\text{int,tog,s}}(t)&= -f(t)g\cos{(\omega_{\text{s}}t+\xi)}\left[\cos{(2\widetilde{\nu}(t))}\sigma_{z}\right.\\
&+\sin{(2\widetilde{\nu}(t))}\left(\e^{i\int_{t_0}^{t}f(t^{\prime})\widetilde{\Omega}_{\text{eff}}(t^{\prime})d t^{\prime}}\sigma_{+}+\text{H. c.}\right)].\notag
\end{align}
We assume that $\omega_{\text{s}}\ll \widetilde{\Omega}_{\text{eff}}(t)$ and that the adiabatic approximation is valid, i.e., $|\widetilde{\nu}^{\prime}(t)|\ll\widetilde{\Omega}_{\text{eff}}(t)$, so we can neglect the fast oscillating second term and obtain.
\begin{equation}\label{Eq:H_ad_sense_interaction_tog_approx_appendix}
H_{\text{int,tog,s}}(t)= -\widetilde{f}(t)g\cos{(\omega_{\text{s}}t+\xi)}\sigma_{z},
\end{equation}
where the modulation function $\widetilde{f}(t)= f(t)\cos{(2\widetilde{\nu}(t))}$.
We note that the modulation function $\widetilde{f}(t)$ would stay the same if the mixing angle changes suddenly
by $\Delta\nu=\pm\pi/2$ between two RAP pulses because the function $f(t)$ and the element $\cos{(2\widetilde{\nu}(t))}$ change their signs simultaneously then. Thus, the modulation function $\widetilde{f}(t)$ is affected only by adiabatic changes in the mixing angle during the RAP pulses (see Fig. 1 in the main text).
%
Next, we consider two approaches for sensing, using adiabatic coherent control.
%

\subsection{Adiabatic evolution sensing}

We first consider the case when the evolution is adiabatic during the whole interaction without sudden changes in the mixing angle. For example, this will be the case if the mixing angle stays constant or changes adiabatically from $\pi/2$ to $0$, then back, etc. As the evolution is adiabatic during the whole interaction, there will be no population changes in the adiabatic basis. Thus, $f(t)=-1$ during the whole interaction and the modulation function will be given by $\widetilde{f}(t)= -\cos{(2\widetilde{\nu}(t))}$. Then, the Hamiltonian in the interaction, toggling basis takes the form
\begin{equation}\label{Eq:H_ad_sense_interaction_tog_approx_CW}
H_{\text{int,tog,s}}(t)= \cos{(2\widetilde{\nu}(t))}g\cos{(\omega_{\text{s}}t+\xi)}\sigma_{z},
\end{equation}
If the mixing angle $\widetilde{\nu}(t)$ stays constant, e.g., if we apply a driving field with a constant Rabi frequency and detuning, the effect of the sensed signal will be cancelled. We note that one can do AC sensing with a simple continuous drive but this requires $\omega_{\text{s}}= \widetilde{\Omega}_{\text{eff}}(t)$ \cite{CaiNJP2012,CohenFP2017,AharonPRL2019} and we consider the case when $\omega_{\text{s}}\ll \widetilde{\Omega}_{\text{eff}}(t)$ in this work. However, if the mixing angle $2\widetilde{\nu}(t)$ changes with a rate, which corresponds to $\pi/\omega_{\text{s}}$, we will be able to sense the signal. For example, a maximum contrast is achieved when the modulation function $\cos{(2\widetilde{\nu}(t))}$ changes its sign 
at the time when $\cos{(\omega_{\text{s}}t+\xi)}$ does this (see Fig. 1 in the main text). If the RAP transition time is very short, i.e., $T_{\text{tr}}\ll \pi/\omega_{\text{s}}$, where $T_{\text{tr}}=2\widetilde{\Omega}(t_{c})/\widetilde{\Delta}^{\prime}(t_{c})$ (see Appendix, sec. \ref{Subsection:RAP_transition_time}) and $t_{c}$ is the time of level crossing in the bare basis, the modulation function can be considered approximately equal to a step function.
Then, the Hamiltonian in Eq. \eqref{Eq:H_effective_sense} can be approximated by
\begin{align}\label{Eq:H_effective_sense}
H_{\text{int,s}}(t)&\approx -g|\cos{(\omega_{\text{s}}t)}|\sigma_{z}
\end{align}
where we assumed that $\xi=0$ for maximum contrast and $\nu(t_0)=\pi/2$ without loss of generality.
As a result of the signal, the sensing qubit will accumulate a phase $2\eta(t)$ in the interaction, toggling basis similarly to standard pulsed DD with instantaneous resonant $\pi$ pulses.
The phase is proportional to $g$ and takes the form (we assume $g\ll \omega_{\text{s}}$)
\begin{equation}\label{eta_t}
\eta(t)\equiv\int_{0}^{t}g|\cos{(\omega_{\text{s}} t^{\prime})}| d t^{\prime}\approx\frac{2}{\pi}g t
\end{equation}
and the effective propagator in this basis is
\begin{equation}
U_{\text{int,s}}(t,t_0)=\cos{\eta(t)}\sigma_0+i\sin{\eta(t)}\sigma_z
\end{equation}
Thus, the sensing qubit accumulates a phase and performs Ramsey oscillations in this basis, similarly to standard pulsed DD.

However, we note that this method for adiabatic sensing is not optimally robust.
For example, if we apply a field with a constant drive and change the target detuning $\Delta(t)$ adiabatically from positive to negative and vice versa at a rate $\pi/\omega_{\text{s}}$, the method will suffer from noise in the effective Rabi frequency $\widetilde{\Omega}_{\text{eff}}(t)$, which defines the basis of the Hamiltonian in Eq. \eqref{Eq:H_effective_sense}. 
This can be seen directly if one considers the effective propagator in the adiabatic basis, which takes the form
\begin{align}
U_{\text{ad,s}}(t,t_0)&=\cos{\left(\frac{\widetilde{\Phi}(t)}{2}+\eta(t)\right)}\sigma_0+
\sin{\left(\frac{\widetilde{\Phi}(t)}{2}+\eta(t)\right)}\sigma_z\notag\\
&=\left[\begin{array}{cc} \e^{i\left(\frac{\widetilde{\Phi}(t)}{2}+\eta(t)\right)} & 0 \\
0 & \e^{-i\left(\frac{\widetilde{\Phi}(t)}{2}+\eta(t)\right)} \end{array} \right].
\end{align}
where the phase $\widetilde{\Phi}(t)=\int_{t_0}^{t}\Omega_{\text{eff}}(t^{\prime}) d t^{\prime}$ depends on noise terms, which will cause the dephasing. We note that the change of the mixing angle reduces the effect of this noise partially. Specifically, if we 
assume that the frequency noise is characterized by $\Delta_{\epsilon}(t)=\Delta_{\epsilon}>0$ and the target detuning $\Delta(t)$ changes from a very low negative value to a very high positive one, we can obtain $\widetilde{\Delta}(t)=\Delta(t)-\Delta_{\epsilon}(t)$ and the effective Rabi frequency $\widetilde{\Omega}_{\text{eff}}(t)=\sqrt{\widetilde{\Delta}(t)^2+\widetilde{\Omega}(t)^2}$ during a RAP pulse by
\begin{subequations}
\begin{align}
\widetilde{\Omega}_{\text{eff}}(t)&\to|\widetilde{\Delta}(t)|=|\Delta(t)|+\Delta_{\epsilon},~\widetilde{\nu}(t)\to\pi/2,\\
\widetilde{\Omega}_{\text{eff}}(t)&\to|\widetilde{\Omega}(t)|,~\widetilde{\nu}(t)\to\pi/4,\\
\widetilde{\Omega}_{\text{eff}}(t)&\to|\widetilde{\Delta}(t)|=|\Delta(t)|-\Delta_{\epsilon},~\widetilde{\nu}(t)\to 0.
\end{align}
\end{subequations}
Thus, the detuning noise due to $\Delta_{\epsilon}$ can be compensated if the time period when $\widetilde{\nu}(t)\to\pi/2$ is equal to the one when $\widetilde{\nu}(t)\to 0$. However, the accumulated phase $\widetilde{\Phi}(t)$ remains susceptible to amplitude noise and higher order frequency noise terms when the mixing angle is changing. Additionally, even in the noiseless case, the dynamic phase due to $\Delta(t)$ and $\Omega(t)$ is not zero and should be taken into account when performing measurements in the bare basis.

\subsection{Sensing by sequences of RAP pulses}

We consider now an improved protocol when we apply sequences of RAP pulses for sensing. Then, the sudden changes in the mixing angle between RAP pulses cause flips of the states in the adiabatic basis, which nullify the dynamic phase and its noise as long as the they occur frequently enough. We consider again the Hamiltonian in the interaction, toggling basis, as defined in Eq. \eqref{Eq:H_ad_sense_interaction_tog_approx}.
\begin{equation}
H_{\text{int,tog,s}}(t)= -\widetilde{f}(t)g\cos{(\omega_{\text{s}}t+\xi)}\sigma_{z},
\end{equation}
where the modulation function $\widetilde{f}(t)=f(t)\cos{(2\widetilde{\nu}(t))}$. We note that we assume that $\widetilde{\nu}(t))$ changes adiabatically from $\pi/2$ to $0$ during every RAP pulse and then instantaneously from $0$ to $\pi/2$ between the pulses (or vice versa for both changes).
We already noted that $\widetilde{f}(t)$ is not affected by sudden changes in the mixing angle when $\Delta\nu=\pm\pi/2$. Thus, the modulation function of the sensed field will be the same as in the case of adiabatic evolution without such changes. Then, if a RAP pulse duration corresponds to $\pi/\omega_{\text{s}}$, we will again be able to sense the signal. Similarly to the case of adiabatic evolution sensing, a maximum contrast is achieved when the modulation function $\widetilde{f}(t)$ changes its sign 
at the time when $\cos{(\omega_{\text{s}}t+\xi)}$ does this (see Fig. 1 in the main text).
%
%

The main difference from continuous RAP sensing is that the interaction, toggling basis itself is much more robust to frequency and amplitude noise as it is defined with respect to $f(t)\Omega_{\text{eff}}(t)$ and $f(t)$ changes its sign during every subsequent RAP pulse. Explicitly, the effective propagator in the adiabatic basis takes the form
\begin{align}
U_{\text{ad,s}}(t,t_0)&=\left[\begin{array}{cc} \e^{i\left(\frac{\widetilde{\Phi}_{c}(t)}{2}+\eta(t)\right)} & 0 \\
0 & \e^{-i\left(\frac{\widetilde{\Phi}_{c}(t)}{2}+\eta(t)\right)} \end{array} \right],
\end{align}
where $\widetilde{\Phi}_{c}(t)=\int_{t_0}^{t}f(t^{\prime})\Omega_{\text{eff}}(t^{\prime}) d t^{\prime}$. It is evident that $\widetilde{\Phi}_{c}(t)=0$ and the accumulated dynamic phase, including the effect of the frequency and amplitude noise, is compensated after every second RAP pulse as long as the correlation time of the noise is long in comparison to the duration of two RAP pulses. Furthermore, the phase evolution in the adiabatic, toggling basis can then be observed stroboscopically directly in the bare basis after every second RAP pulse. We note that as the instantaneous changes in the mixing angle do not affect the modulation function $\widetilde{f}(t)$ (but only $f(t)$), the RAP pulses can also be truncated and separated by free evolution time $\tau$ (see Fig. 1 in the main text). Then, the sensing condition becomes $T_{\text{pulse}}+\tau=\pi/\omega_{\text{s}}$.
However, unless experimental limitations require such truncation, it is usually beneficial to use longer RAP pulses and $\tau=0$ as this improves adiabaticity.

Finally, we note that while the instantaneous changes in the mixing angle play the role of instantaneous $\pi$ pulses around the $y$ axis in the adiabatic basis, they do not compensate errors due to non-adiabatic couplings. The reason is that the Hamiltonian term due to the latter 
is proportional to $\sim \nu^{\prime}(t)\sigma_{y}$ and commutes with the Hamiltonian during the sudden change of the mixing angle. Additionally, the changes in the mixing angle during/between RAP pulses might differ from $\pi/2$. In order to compensate for these imperfections, we apply phased sequences of RAP pulses and use their relative phases additional control parameters to improve the fidelity of the process, as discussed in sec. \ref{Subsection:Phased_RAP}.

\section{Detailed comparison of RAP sensing and sensing with rectangular pulses}\label{Section:RAP_simulation_comparison}

Sensing by sequences of RAP pulses allows to obtain an improved contrast in comparison to sensing with rectangular $\pi$ pulses. This is due to the greater bandwidth and robustness to amplitude errors of the RAP, e.g., for systems with large inhomogeneous broadening.
Specifically, it can be shown that the obtained contrast in sensing experiments with a Hahn echo with an imperfect pulse is proportional to the transition probability $\sim p$ of the latter \cite{DegenRMP2017}. The relation is more complicated with longer phased sequences, e.g., XY8 (see Appendix, sec. \ref{Section:Robust_RAP_sequences}), but higher $p$ in general leads to improved contrast and coherence times.

Standard rectangular $\pi$ pulses require a peak Rabi frequency of $\Omega_0 \gg \Delta_{\text{inh}}$ in order to have sufficient bandwidth to cover the full width of the inhomogeneous broadening. Specifically, the error in the transition probability is given by
\begin{equation}
\epsilon_{\text{rect}}=1-\frac{\Omega_0^2}{\Omega_{\text{eff}}^2}\sin{(\Omega_{\text{eff}}T_{\text{pulse}}/2)}\sim \frac{\Delta_{\text{inh}}^2}{\Omega_{\text{eff}}^2}\approx \frac{\Delta_{\text{inh}}^2}{\Omega_{0}^2},
\end{equation}
where $\Omega_0$ is the Rabi frequency, $T_{\text{pulse}}$ is the pulse duration, $\Delta_{\text{inh}}$ is the detuning of the applied field from the frequency of the sensor qubit, e.g., due to inhomogeneous broadening. Finally,  $\Omega_{\text{eff}}=\sqrt{\Omega_0^2+\Delta_{\text{inh}}^2}$ is the effective Rabi frequency, with the last approximations valid for small detunings. One can see that the error in the transition probability with rectangular pulses can be significant when $\Delta_{\text{inh}}$ is large in comparison to $\Omega_0$ or in case of variation of the Rabi frequency, so that the effective pulse area $\Omega_{\text{eff}}T_{\text{pulse}}\ne \pi$.

RAP pulses are robust to frequency and amplitude variation and their transition probability depends on the particular pulse shape and time dependence of the detuning \cite{Vitanov01ARPC}. One can obtain an approximate estimate of the transition probability error by considering the probability for non-adiabatic transitions if we assume that the mixing angle changes from $\pi/2$ to $0$ during a pulse. The transition probability in the adiabatic basis is determined from the Hamiltonian in Eq. \eqref{Eq:H_ad_no_approx_appendix} and can be approximated by
\begin{equation}
\epsilon_{\text{RAP}}
\sim \left(\frac{\dot{\Delta}(t_{\text{c}})}{\Omega(t_{\text{c}})^2}\right)^2 =\left(\frac{R/2}{\Omega_0^2 T}\right)^2\sim\left(\frac{R}{\Omega_0^2 T_{\text{pulse}}}\right)^2,
\end{equation}
where $t_{\text{c}}$ is the time of the level crossing, and the second equality is valid for the Allen-Eberly (AE) model (see below) with $R$ - the target chirp range, $T\sim T_{\text{pulse}}$ - the characteristic time of the chirped pulse. As another example, in the case of widely used pulse with a constant Rabi frequency and a linear chirp i.e., the standard Landau-Zener-Stückelberg-Majorana (LZSM) \cite{LandauZener1932}, the error in the transition probability in the limit of very long pulse duration is given by $\epsilon_{\text{RAP}}=\exp{\left(-\frac{\pi \Omega_0^2 T_{\text{pulse}}}{2R}\right)}$ \cite{Kyoseva06PRA} and we again obtain a dependence on the parameter $R/(\Omega_0^2 T_{\text{pulse}})$.
The sensing condition with chirped pulses requires $T_{\text{pulse}}\sim\pi/\omega_s$ and $R\sim \Delta_{\text{inh}}$ in order for the chirp range to cover the inhomogeneous broadening, so one can obtain
\begin{equation}
\epsilon_{\text{RAP}}\sim \frac{R^2}{\Omega_{0}^2}\frac{\omega_{s}^2}{\Omega_0^2}\sim \frac{\Delta_{\text{inh}}^2}{\Omega_{0}^2}\frac{\omega_{s}^2}{\Omega_0^2}\approx \epsilon_{\text{rect}}\frac{\omega_{s}^2}{\Omega_0^2}.
\end{equation}
Thus, the error in the transition probability is lowered by $\sim \omega_{s}^2/\Omega_0^2$.
As a result, the RAP sensing protocol would improve performance significantly in comparison to rectangular pulses when $\Omega_{0}<\Delta_{\text{inh}}$ and $\omega_{s}\ll\Omega_{0}$. It is also less sensitive to variation in the effective pulse area in comparison to the rectangular pulses, so it would also be applicable in the case of Rabi frequency inhomogeneity.

Next, we discuss the AC signal frequency range, which can be sensed with RAP pulses. The latter are typically longer than the standard rectangular pulses, so they are preferable for sensing of low frequency AC signals. The upper limit of the sensed frequency can be determined from the estimated error in the transition probability, e.g., of the LZSM model $\epsilon_{\text{RAP}}=\exp{\left(-\frac{\pi \Omega_0^2 T_{\text{pulse}}}{2R}\right)}\ll 1$ \cite{Kyoseva06PRA}, which requires
\begin{equation}
\omega_{s}\ll \frac{\pi^2\Omega_{0}^2}{2R}\sim \frac{\pi^2\Omega_{0}^2}{4\Delta_{\text{inh}}},
\end{equation}
where we used that $\omega_{s}=\pi/T_{\text{pulse}}$. We note that this limit can increase significantly by using other pulse shapes or phased sequences of chirped pulses that improve the fidelity of the process, e.g.,the fidelity error of the XY8 sequence with chirped pulses is $\sim \epsilon_{\text{RAP}}^3 \ll \epsilon_{\text{RAP}}$ (see Appendix, sec. \ref{Section:Robust_RAP_sequences}).

The lower limit of the sensed frequency is determined by the $T_2$ time of DD with ideal, instantaneous $\pi$ pulses with a pulse separation $\pi/\omega_{s}$, e.g., due to homogeneous broadening. For example, if we assume that the homogeneous broadening noise spectrum is given by the Lorentzian $S(\omega)=\frac{b^2}{\pi}\frac{1/\widetilde{\tau}}{(1/\widetilde{\tau})^2+\omega^2}$, where $\widetilde{\tau}$ is the correlation time of the environment and $b$ 
is the bath coupling strength 
(see Appendix, sec. \ref{Section:Numerics}), the decay of the signal after a single pulse can be approximated by $\sim \exp{\left(-\frac{b^2 T_{\text{pulse}}^3}{12\widetilde{\tau}}\right)}$ \cite{deLangeScience2010}. Thus, we require
\begin{equation}
\omega_{s}\gg \pi \left(\frac{b^2}{12\widetilde{\tau}}\right)^{1/3}.
\end{equation}

Thus, the sensing frequency range of RAP sequences is determined by the repetition rate of the RAP pulses. As they are typically long to ensure adiabaticity, the resulting slower repetition rate (in comparison to rectangular pulses) makes the protocol sensitive to high frequency noise. Additionally, when the condition that the RAP transition time $T_{\text{tr}}\ll\omega_s/\pi$ is not fulfilled, there can be a slight shift in the amplitude of the detected AC field but it is straightforward to be taken into account. Finally, when the inhomogeneous broadening is large, the transitions of the different sensor atoms happen at different times, i.e., not at the moment when the sensed field is zero, which can lead to a slightly lower contrast.

We also note that in some cases the amplitude and frequency inhomogeneities can also affect the preparation and readout efficiency of the sensing protocol, e.g., leading to a lower contrast. For example, $\pi/2$ rectangular pulses are typically applied to prepare the system in the $|1_{x,y}\rangle$ state and read it out after the sensing experiment. However, one cannot prepare efficiently all atoms when the inhomogeneous broadening is much greater than the bandwidth of the simple $\pi/2$ pulse. One way to address this problem is to use adiabatic half passage pulses for preparation and readout (see Appendix, section \ref{Section:Preparation} for details). Various other techniques can also be applied to improve the preparation and readout efficiency even further, e.g., robust composite $\pi/2$ pulses \cite{Levitt97Review}, adiabatic robust pulses \cite{Tannus97AdiabBook,GarwoodJMR1991,ZlatanovPRA2020}, single-shot shaped pulses \cite{NdongJPB2015,Van-DammePRA2017}, pulses designed by optimal control \cite{SkinnerJMR2003,HaberlePRL2013,BraunNJP2014,ScheuerNJP2014,DoldeNatComm2014,NobauerPRL2015}.

Adiabaticity requirements can be relaxed by the application of phased RAP pulses, similarly to the ones used in this work. Furthermore, the pulse repetition rate is usually determined by the sensed (Larmor) frequency, which cannot be increased in some cases. Finally, the variation in transition times for the different sensor atoms can be used to design more complex filter functions for sensing and dynamical decoupling. Thus, sensing with phased RAP pulses can provide significant advantages in a broad range of applications.

\begin{figure*}[t!]
\includegraphics[width=\textwidth]{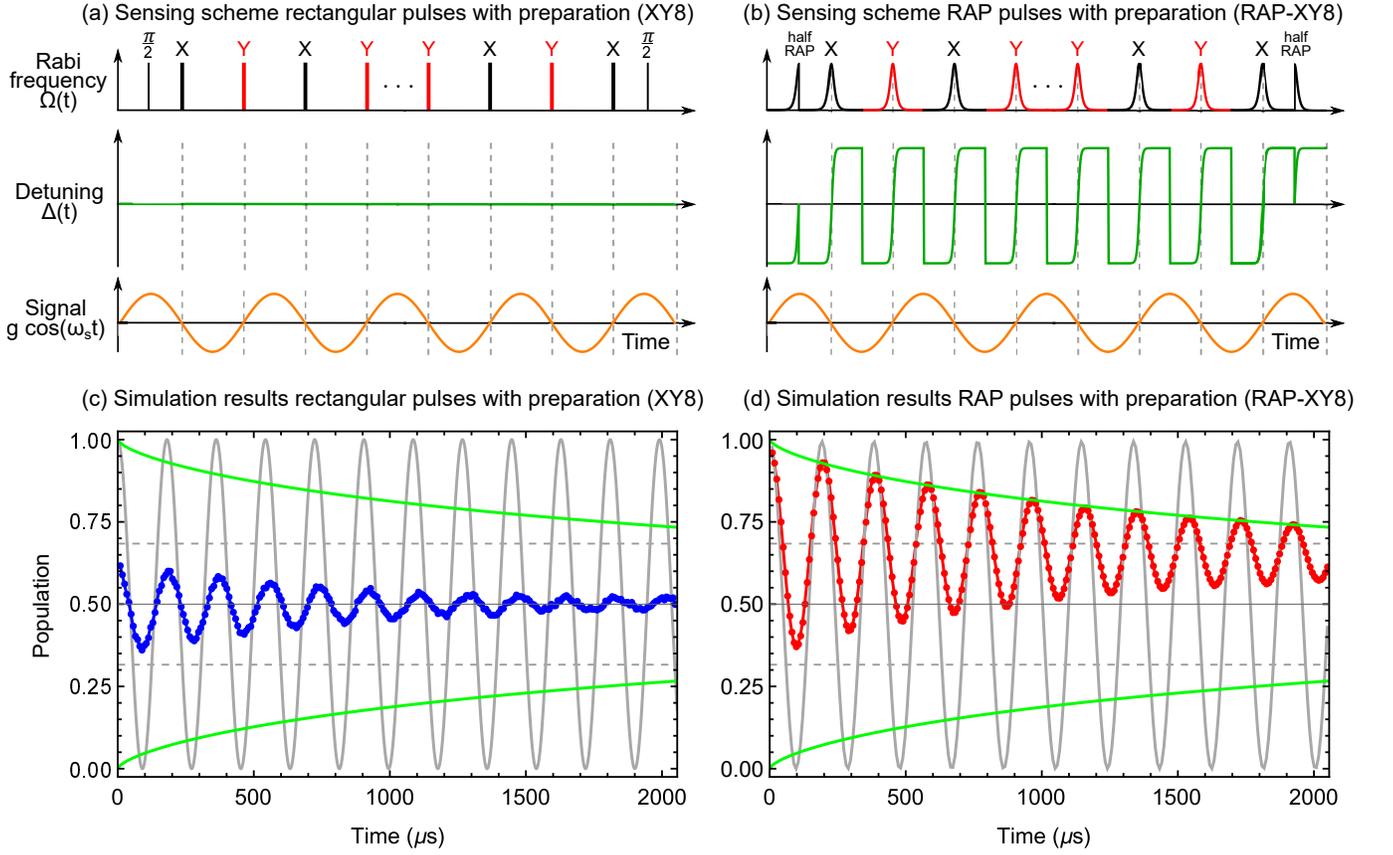}
\caption{(color online)
(a) Scheme for sensing with preparation and readout for (a) the standard XY8 sequence of rectangular pulses and (b) the RAP-XY8 sequence of phase shifted, chirped, adiabatic pulses. In both schemes we assume that the atoms are initially in the ground state $|1_{z}\rangle$, so we need to prepare the system in a coherent superposition state, e.g., by a $\pi/2$ pulse, perform sensing and then readout. Corresponding numerical simulations of the population of the excited state $|1_{-z}\rangle$ in the bare basis for quantum sensing, observed stroboscopically directly in the bare basis
 at time intervals of $8~\mu$s with dynamical decoupling with (c) XY8 with rectangular pulses,
 and (d) RAP-XY8 with chirped pulses.
 The experimental parameters are the same as in Fig. 3 in the main text with the only difference that we apply $\pi/2$ and half-RAP pulses for preparation and readout, respectively.
 The light gray curve shows the respective theoretical evolution in an ideal system without inhomogeneous broadening, frequency and amplitude noise. The red curve shows the evolution with inhomogeneous broadening, frequency and amplitude noise. We note that the slight delay in the ideal, theoretical curve with RAP pulses from $p=\cos(\eta(t))^2$, defined in Eq. \eqref{eta_t}, is expected and due mainly to the non-instantaneous transition time, which is taken into account in the simulation.
 Both the coherence time and the contrast are much higher with the RAP-XY8 protocol.}
\label{Fig:rect_RAP_comparison_with_preparation}
\end{figure*}

\section{Robust preparation and readout}\label{Section:Preparation}

As noted in the main text, applying RAP pulses for sensing increases significantly the contrast and coherence time in systems with large driving field variation and inhomogeneous broadening. In some cases, these inhomogeneities can also affect the preparation and readout efficiency of the sensing protocol. For example, a simple $\pi/2$ pulse cannot prepare efficiently all atoms in an ensemble when the inhomogeneous broadening is much greater than the pulse bandwidth.

Various techniques can be applied to improve the efficiency and robustness of preparation and readout, e.g., one can apply robust composite $\pi/2$ pulses \cite{Levitt97Review}, adiabatic robust pulses \cite{Tannus97AdiabBook,GarwoodJMR1991,ZlatanovPRA2020}, single-shot shaped pulses \cite{NdongJPB2015,Van-DammePRA2017}, pulses designed by optimal control \cite{SkinnerJMR2003,HaberlePRL2013,BraunNJP2014,ScheuerNJP2014,DoldeNatComm2014,NobauerPRL2015}. Figure \ref{Fig:rect_RAP_comparison_with_preparation} shows an example for sensing with RAP pulses with a robust preparation and readout where we replace the simple $\pi/2$ pulses in the standard sensing scheme with adiabatic half passage pulses (half-RAP) pulses. We note that although the preparation and readout efficiency is better with half-RAP than with rectangular pulses, it still reduces contrast slightly in comparison to the case with perfect preparation and readout in Fig. 3 in the main text. This is expected from theory as the inhomogeneous broadening is much larger than the Rabi frequency, so not all atoms are prepared in equal coherent superposition states. Nevertheless, the RAP-XY8 scheme has both better contrast and longer coherence times than the standard XY-8 sensing with rectangular pulses. We note that in standard pulsed DD experiments it is also common to apply measurements with a $3\pi/2$ readout pulse and take the difference (contrast) from the measured population with a standard $\pi/2$ readout pulse. The robust RAP sensing alternative of the $3\pi/2$ readout pulse can be achieved by adding and additional RAP pulse at the end of the sequence in Fig. \ref{Fig:rect_RAP_comparison_with_preparation}. Finally we note that the preparation and readout protocol can be improved further, e.g., by some of the techniques mentioned above, but this goes beyond the scope of this work.

\section{Numerical Simulation}\label{Section:Numerics}

In order to compare sensing with rectangular and RAP pulses, we perform a numerical simulation. The results from the latter are shown in Fig. 3 in the main text and compare the performance of the XY8 and RAP-XY8 protocols in a realistic conditions for sensing in NV centers with large inhomogeneous broadening.
Specifically, we apply dynamical decoupling by sequences of phased RAP pulses in a two-state system with a Hamiltonian in the bare basis
\begin{align}\label{Eq:H_general_RWA}
H_{s}(t)=&-\frac{\widetilde{\Delta}(t)}{2}\sigma_{z}+\frac{\widetilde{\Omega}(t)}{2}\left(\cos{[\phi(t)]}\sigma_{x}+\sin{[\phi(t)]}\sigma_{y}\right)\notag\\
&+g\sigma_{z}\cos{(\omega_{\text{s}}t+\xi)},
\end{align}
where $\widetilde{\Delta}(t)\equiv \Delta(t)-\Delta_{\epsilon}(t)$ is the actual detuning, experienced by a sensor atom, where $\Delta(t)$ is the target detuning and $\Delta_{\epsilon}(t)$ is noise in the transition frequency of the qubit, e.g., due to inhomogeneous broadening or frequency fluctuations. Next, the actual Rabi frequency is $\widetilde{\Omega}(t)=\Omega(t)[1+\epsilon_{\Omega}(t)]$, where $\Omega(t)$ is the target Rabi frequency we want to apply and $\epsilon_{\Omega}(t)$ is an error term, e.g., due to amplitude fluctuations and/or inhomogeneity. Additionally, $\phi(t)$ is a time-dependent phase of the control field, which takes discrete values during each pulse. Finally, $g$ is the amplitude of the oscillating sensed field, $\omega_{\text{s}}$ is its angular frequency and $\xi$ is its initial phase.

First, the target Rabi frequency and detuning of the $k$-th RAP pulse follow the time-dependence of the Allen-Eberly (AE) model \cite{HioePRA1984,Allen-Eberly1987,Kyoseva06PRA}
\begin{subequations}\label{Eq:AE_model_appendix_sim}
\begin{align}
\Omega(t)&=\Omega_0~\sech{\left(\frac{t-t_{c,k}}{T}\right)}\\
\Delta(t)&=\Delta_0 \tanh{\left(\frac{t-t_{c,k}}{T}\right)},
\end{align}
\end{subequations}
for $t\in \left[t_{c,k}-\frac{T_{\text{pulse}}}{2},t_{c,k}+\frac{T_{\text{pulse}}}{2}\right]$, where $t_{c,k}$ is the center of the $k$-th pulse, $T$ is its characteristic time, and $T_{\text{pulse}}$ is the RAP pulse duration. The peak Rabi frequency and detuning are, respectively, $\Omega_0$ and $\Delta_0=R/2$ with $R$ the target chirp range. We note that one can apply chirped pulses with other shapes and detunings, e.g., the standard Landau-Zener-Stückelberg-Majorana model with a constant drive and a linear chirp \cite{LandauZener1932}. We choose the AE model due to its excellent adiabaticity with respect to peak Rabi frequency and chirp range (see Appendix, sec. \ref{Subsection:AE}), allowing for high flexibility of applications.


We assume detuning noise $\Delta_{\epsilon}(t)$ and uncorrelated amplitude fluctuation $\epsilon_{\Omega}(t)$ of the driving field. The parameters of the noise have the characteristics for typical experiments in NV centers, as described in \cite{CaiNJP2012,AharonNJP2016}. Specifically, we assume that the magnetic noise has a constant and a dynamic component $\Delta_{\epsilon}(t)=\Delta_{\epsilon,c}+\Delta_{\epsilon,d}(t)$. The constant component  $\Delta_{\epsilon,c}$ follows a Gaussian distribution with a zero expectation value and a FWHM of $2\pi~26.5$ MHz ($T_2^{\star}=20$ ns). The dynamic component $\Delta_{\epsilon,d}(t)$ has a Lorentzian power spectrum $S(\omega)=\frac{b^2}{\pi}\frac{1/\widetilde{\tau}}{(1/\widetilde{\tau})^2+\omega^2}$, where $\widetilde{\tau}$ is the correlation time of the environment and $b=\sqrt{c \widetilde{\tau}/2}=2\pi~50$ kHz 
is the bath coupling strength with $c$ the diffusion constant.
The component $\Delta_{\epsilon,d}(t)$
is modelled as an Ornstein-Uhlenbeck (OU) process \cite{UhlenbeckRMP1945,GillespieAJP1996} with a zero expectation value $\langle \Delta_{\epsilon,d}(t)\rangle = 0$, correlation function $\langle \Delta_{\epsilon,d}(t)\Delta_{\epsilon,d}(t^{\prime})\rangle =(1/2)c\widetilde{\tau}\exp{(-\gamma|t-t^{\prime}|)}$,
$\widetilde{\tau}=1/\gamma = 20 \mu$s is the correlation time of the noise. The OU process is implemented with an exact algorithm \cite{GillespieAJP1996}
\begin{equation}\label{Eq:OU_noise}
\Delta_{\epsilon,d}(t+\Delta t)=\Delta_{\epsilon,d}(t)\e^{-\frac{\Delta t}{\widetilde{\tau}}}+\widetilde{n}\sqrt{\frac{c\widetilde{\tau}}{2}\left(1-\e^{-\frac{2\Delta t}{\widetilde{\tau}}}\right)},
\end{equation}
where $\widetilde{n}$ is a unit Gaussian random number. 
The driving fluctuations are also modelled by uncorrelated OU processes with the same correlation time $\tau_{\Omega}=500 \mu$s and a relative amplitude error $\epsilon_{\Omega}=0.005$ with the corresponding diffusion constant $c_{\Omega}=2\delta_{\Omega_{i}}^2\Omega_{i}^2/\tau_{\Omega},~i=1,2$.

Then, we calculate numerically the propagator
\begin{equation}
\widetilde{U}_{s}(t,t_0)=\T \exp{\left(-i\int_{t_0}^{t}\widetilde{H}_{s}(t^{\prime}) d t^{\prime}\right)}
\end{equation}
for the particular noise realisation of $\Delta_{\epsilon}(t)$and $\epsilon_{\Omega}(t)$ and the chosen DD sequence. We use a time-discretization with a time step of $0.1$ ns, which is comparable to the resolution of available arbitrary wave-form generators. We note that the OU noise characteristics are not affected by this choice of $\Delta t$, as Eq. \eqref{Eq:OU_noise} is exact.

We then make use of the calculated $\widetilde{U}_{s}(t,t_0)$ and obtain the time evolution of the density matrix
\begin{equation}
\rho(t)=\widetilde{U}_{s}(t,t_0)\rho(t_0)\widetilde{U}^\dagger_{s}(t,t_0),
\end{equation}
where $\rho(t_0)=\rho_{y}\equiv(\sigma_{0}+\sigma_{y})/2$ is the initial density matrix.
We assume in the simulation in Fig. 3 in the main text that $\rho(t_0)=\rho_{y}\equiv(\sigma_{0}+\sigma_{y})/2$, which corresponds to perfect
preparation of the system in the state $|1_{y}\rangle$. We note that the initial state can also be $|1_{x}\rangle$ or any other state, which has components that do not commute with a $\sim\sigma_{z}$ Hamiltonian in order to sense the signal. The expected density matrix $\overline{\rho}(t)$ is calculated by performing the simulation $2500$ times for different noise realizations and averaging the result. The simulation
results in Fig. 3 in the main text
show the average population in state $|1_{y}\rangle$, which is calculated as $P_{1y}(t)=(1/2)+\text{Im}(\overline{\rho}_{21}(t))$.

The simulation in Fig. \ref{Fig:rect_RAP_comparison_with_preparation} assumes
\red
$\rho(t_0)=\rho_{y}\equiv(\sigma_{0}+\sigma_{y})/2$
\black
and takes into account imperfect preparation and readout. We calculate the expected density matrix $\overline{\rho}(t)$ and show the average population in state $|1_{-z}\rangle$, which is determined by $P_{1-z}(t)=\overline{\rho}_{22}(t)$.

\begin{figure}[t!]
\includegraphics[width=0.8\columnwidth]{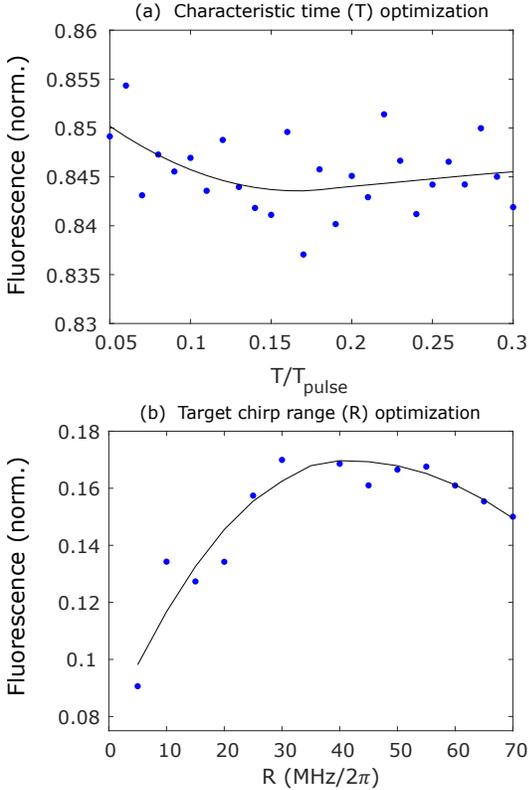}
\caption{(color online)
 RAP experimental optimization for the High Rabi experiment with peak Rabi frequency \(\Omega_{0}= 2\pi\:5\:\) MHz and pulse duration \(T_{\text{pulse}}=11.4\:\mu s\). (a) Fluorescence vs. ratio characteristic time $T/T_{\text{pulse}}$ for a single RAP pulse. The lower the fluorescence the more efficient is the population from $|0\rangle$ to $|1\rangle$ (see Fig. 3(c) in the main text). The chirp range is taken \(R=2\pi\:40\: \) MHz. (b) Optimization of the target chirp range $R$ for a characteristic time \(T/T_{\text{pulse}}=0.17\) for the RAP-XY8 DD sequence, repeated once. 
 }
\label{fig:RAP_optimization}
\end{figure}

\begin{figure*}[t!]
\includegraphics[width=\textwidth]{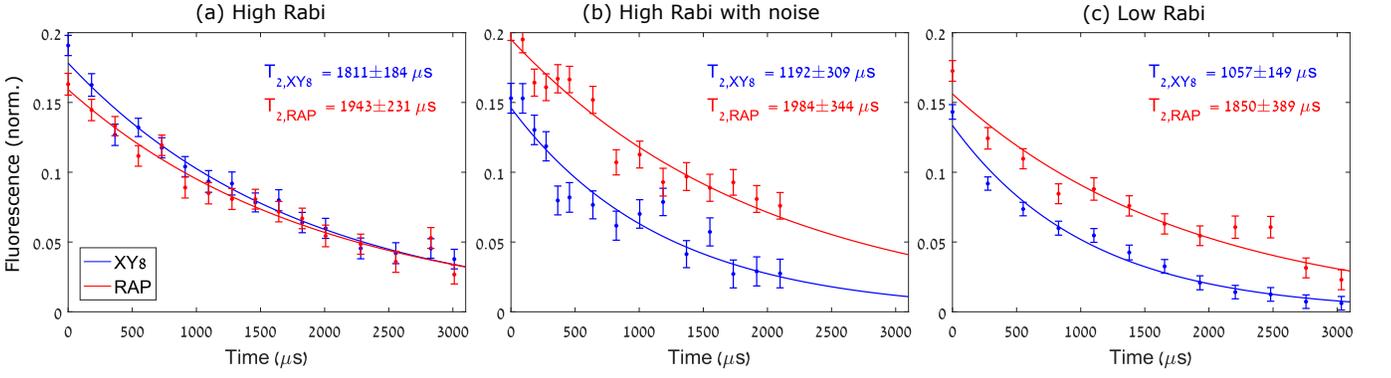}
\caption{(color online)
 Fluorescence vs. interaction time for XY8 and RAP-XY8 without an artificial AC field for (a) High Rabi experiment with a peak Rabi frequency \(\Omega_{0}=2\pi\:5\:\) MHz, \(T_{\text{pulse}}+\tau=11.4\:\mu s\). The rectangular pulses in standard XY8 have a duration \(T_{\text{pulse}}=\pi/\Omega_0=100\:n s\), $\tau=11.3~\mu$s; for RAP-XY8: \(T_{\text{pulse}}=11.4\:\mu s\), $\tau=0$, target chirp range \(R=2\pi\:40\: \) MHz, characteristic time \(T/T_{\text{pulse}}=0.17\). (b) High Rabi with noise: the same pulse parameters as in (a) plus added amplitude noise with \(\sigma=0.2\:\Omega_{0}\). The coherence time \(T_2\) of standard XY8 decreases by \(\sim35\%\) in comparison to the High Rabi case while remaining approximately the same with RAP-XY8. (c) Low Rabi experiment with peak Rabi frequency \(\Omega_{0}=2\pi\:1.7\:\) MHz, \(T_{\text{pulse}}+\tau=34.46\:\mu s\). The rectangular pulses in standard XY8 have a duration \(T_{\text{pulse}}=\pi/\Omega_0=294\:n s\), $\tau=34.166~\mu$s; for RAP-XY8: \(T_{\text{pulse}}=34.46\:\mu s\), $\tau=0$, target chirp range \(R=2\pi\:20\: \) MHz, characteristic time \(T/T_{\text{pulse}}=0.19\).
 The coherence time \(T_2\) of standard XY8 decreases by \(\sim40\%\) in comparison to the High Rabi case while remaining approximately the same with RAP-XY8. All experiments show fluorescence difference from two measurements in which the phase of the last $\pi/2$ or half RAP pulse is $X$ and $-X$.}
\label{fig:Coherence-decay-curves}
\end{figure*}

\begin{figure*}[t!]
\includegraphics[width=0.9\textwidth]{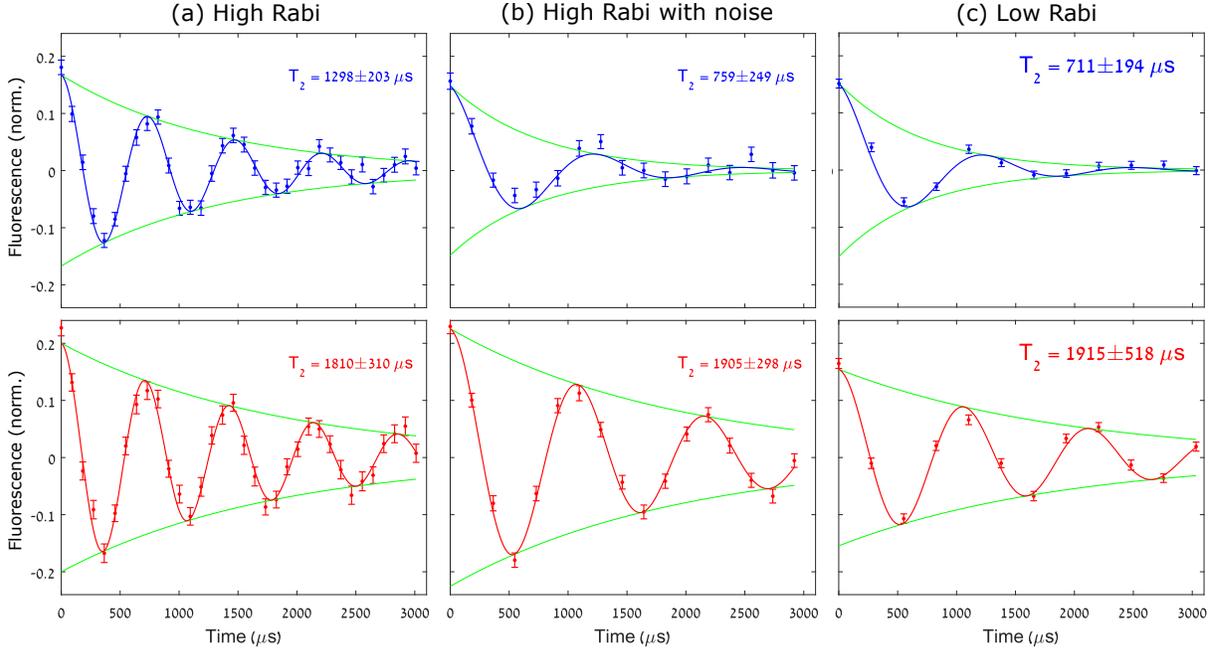}
\caption{(color online)
 Fluorescence vs. interaction time for a quantum sensing experiment with (top) XY8 and (bottom) RAP-XY8 with an artificial AC field with a frequency corresponding to half the pulse repetition rate. The pulse parameters are the same as in Fig. \ref{fig:Coherence-decay-curves}. The magnitude and angular frequency of the artificial sensed field are respectively (a) $78~nT$ and $\omega_s = 2\pi~43.8$ kHz, (b) $52~nT$ and $\omega_s = 2\pi~43.8$ kHz, and (c) $52~nT$ and $\omega_s = 2\pi~14.5$ kHz. The coherence time \(T_2\) of standard XY8 in the High Rabi with noise experiment decreases by \(\sim40\%\) in comparison to the High Rabi case while remaining approximately the same with RAP-XY8.
 The corresponding drop of \(T_2\) of standard XY8 is \(\sim45\%\) for the Low Rabi case while again remaining approximately the same with RAP-XY8. All experiments show fluorescence difference from two measurements in which the phase of the last $\pi/2$ or half RAP pulse is $X$ and $-X$.}
\label{fig:Sensing_all}
\end{figure*}

\begin{figure*}[t!]
\includegraphics[width=0.95\textwidth]{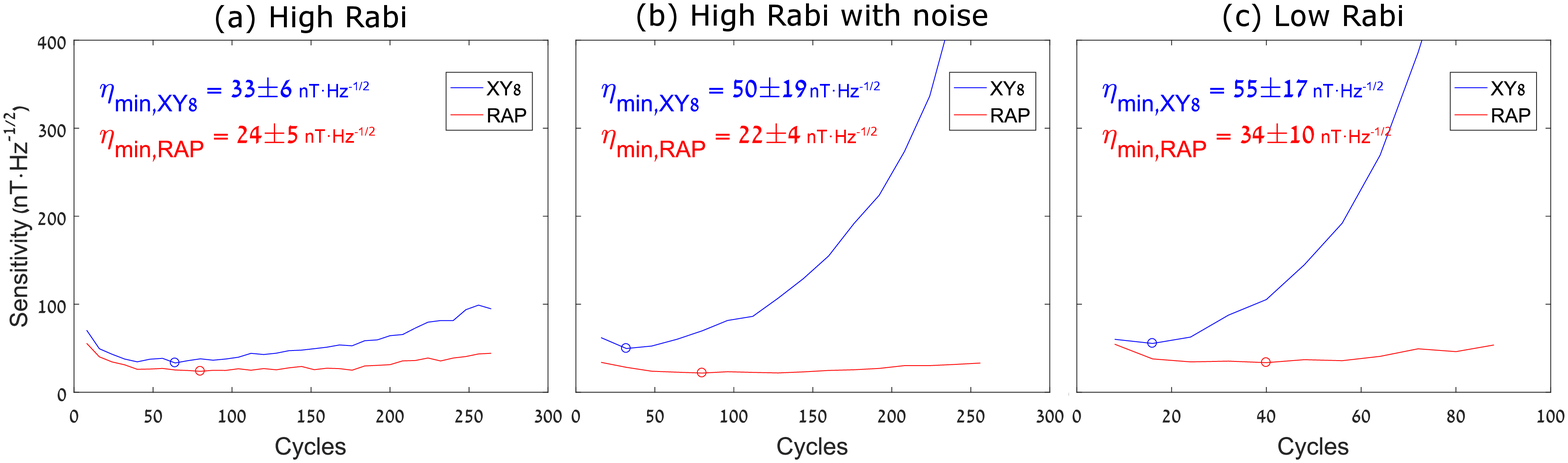}
\caption{(color online)
Sensitivity vs. number of free evolution-pulse-free evolution cycles for pulse parameters and characteristics of the artificial AC field as in Fig. \ref{fig:Sensing_all}. }
\label{fig:All_sensitivity}
\end{figure*}

\section{Additional experimental results}\label{Section:SM_Experimental_results}

In the following section, we give more details of the performed experiments for RAP sensing.
In a first experiment, we optimized experimentally the RAP pulse parameters for the Allen-Eberly model \cite{Allen-Eberly1987}. The procedure is also applicable for the optimization of RAP pulses with other shapes. An example for this optimization for the High Rabi experiment is shown in Fig. \ref{fig:RAP_optimization}, where we optimize the characteristic time $T$ and the target chirp range $R$. In this specific example, we first fix the peak Rabi frequency \(\Omega_{0}= 2\pi\:5\:\) MHz, which is usually determined by the experimental limitations. We assume RAP pulse duration \(T_{\text{pulse}}=11.4\:\mu s\), which is determined by the frequency of the sensed field in a subsequent experiment. As a next step, we optimize the RAP chacteristic time $T$, which determines the extent of truncation of the RAP pulse. We initialize the system in the $|0\rangle$ state and perform population inversion by a RAP pulse, measuring the observed fluorescence (see Fig. 3c in the main text), which is correlated to the population distribution between the $|0\rangle$ and $|1\rangle$ states at the end of the interaction. We choose a chirp range, which is expected to perform well in theory and scan $T$ while keeping $T_{\text{pulse}}$ constant. In this way, we obtain an optimum ratio \(T/T_{\text{pulse}}=0.17\) (see Fig. \ref{fig:RAP_optimization}(a)). Next, we use this \(T/T_{\text{pulse}}\) and optimize the target chirp range $R$ by performing DD by RAP-XY8, repeated once, in order to maximize the DD efficiency. Figure \ref{fig:RAP_optimization}(b) shows that we obtain the optimum \(R=2\pi\:40\: \) MHz. We note that in all DD experiments, including for the $R$ optimization, we obtain the difference in fluorescence from two measurements in which the phase of the last half RAP pulse is $X$ and $-X$. As we plot a fluorescence difference in Fig. \ref{fig:RAP_optimization}(b), the values are expectedly lower in comparison to Fig. \ref{fig:RAP_optimization}(a), where we plot the fluorescence after a single measurement. The optimization can also be performed by a 2D scan of these parameters but our results show that slight variations of $T$ and $R$ do not affect the results significantly, so we use this simplified optimization.

Next, we carried out three sets of experiments:
(i) at high Rabi frequency ($2 \pi~5$ MHz), which we term High Rabi in all figures; (ii) at high Rabi frequency with artificially added amplitude error for each sequence repetition, which follows a random Gaussian distribution with a zero mean and a width of \(0.2~\Omega_{0}\), which we term High Rabi with noise; (iii) at low Rabi frequency ($2 \pi~1.7$ MHz), which we term Low Rabi.
Each set included regular XY8 (``hard'', rectangular pulses) and RAP XY8 sequences where the pulse and detuning shape follow the Allen-Eberly model in Eqs. \eqref{Eq:AE_model_appendix_sim} with the abovementioned values of the Rabi frequency corresponding to the peak Rabi frequency. We performed the measurement twice for each sequence: without an external field to be sensed (a decoherence measurement, see Fig. \ref{fig:Coherence-decay-curves}) and with an applied oscillating magnetic field (acting as the signal to be sensed, see Fig. \ref{fig:Sensing_all}).


In the first series of experiments, we measured the coherence time ($T_{2}$) of the DD sequences without applying an artificial oscillating magnetic field (see Fig. \ref{fig:Coherence-decay-curves}).
Specifically, we measured the change in fluorescence, which varies depending on the population of the electron spin of the NV center. This allowed us to measure the decoherence of the electron spin as a function of time, while applying DD with pulses at a constant repetition rate (so, that the number of pulses changes with time).
As shown in Fig. \ref{fig:Coherence-decay-curves}(a) the coherence time with the standard XY8 sequence with rectangular pulses is around $1.8$ ms with strong, rectangular pulses. As expected from theory, the standard XY8 DD protocol does not perform well when we introduce amplitude noise and the coherence time drops to about $1.2$ ms in Fig. \ref{fig:Coherence-decay-curves}(b). Similarly, in Fig. \ref{fig:Coherence-decay-curves}(c), we observe a drop in the coherence time to about 1 ms as we apply a low Rabi frequency of \(\Omega_{0}=2\pi\:1.7\:\) MHz, which is less than the FWHM of the inhomogeneous broadening of $ 2 \pi \times (2.1 \pm 0.1)$ MHz, so the rectangular DD pulses cannot cover efficiently the full bandwidth of transition frequencies of the NVs in the ensemble.
The coherence times improve with the RAP-XY8 sequence and reach $T_2 \approx 1.94$ ms. In addition, RAP XY8 is robust to amplitude and frequency variation as the coherence times remain approximately the same for all experimental variants. Specifically, they are insensitive to added amplitude noise and perform well even when the Rabi frequency is lower than the inhomogeneous broadening and the standard approach does not work efficiently (see Figs. \ref{fig:Coherence-decay-curves}(b-c)). 

In the second series of experiments, we performed quantum sensing by adding an external AC magnetic field in order to characterize the magnetic sensitivity of our system under regular control pulses and RAP. We measured the change in fluorescence, which varies depending on the population of the electron spin of the NV center, as a function of time. We also applied pulsed DD to decouple from the noise from the environment. The frequency of the AC field was matched to half DD pulse repetition rate, which was kept constant (such that the number of pulses changes with time). The presence of the external field reduced the coherence time of the NV electron spin by $\sim 30 \%$ using standard XY8 (see Fig. \ref{fig:Sensing_all}(a), top). We attribute this effect to the rotation of the NV electron spin state in the XY plane of the Bloch sphere due to the field, so the system is more sensitive to pulse imperfections. While this effect should be negligible in the small-field limit \cite{PhamPRB2012}, it does not adversely affect RAP sequences even for larger fields (Fig. \ref{fig:Sensing_all}(a), bottom).
As expected from theory, the standard XY8 DD protocol does not perform well when we introduce amplitude noise and the coherence time drops by about 40\% in Fig. \ref{fig:Sensing_all}(b), top. Similarly, in Fig. \ref{fig:Sensing_all}(c), top, we observe a drop in the coherence time by about 45\% in comparison to the High Rabi case as we apply a low Rabi frequency of \(\Omega_{0}=2\pi\:1.7\:\) MHz, which is less than the FWHM of the inhomogeneous broadening of $ 2 \pi \times (2.1 \pm 0.1)$ MHz. Thus, the rectangular DD pulses cannot cover efficiently the full bandwidth of transition frequencies of the NVs in the ensemble and the coherence time drops.
In contrast to the standard XY8 sensing, the coherence times with the RAP-XY8 sequence are not affected by the presence of an artificial field (see Fig. \ref{fig:Sensing_all}(a),bottom). They are also insensitive to the added amplitude noise and perform well even when the Rabi frequency is lower than the inhomogeneous broadening and the standard approach does not work efficiently (see Fig. \ref{fig:Sensing_all}(b-c),bottom). Thus, in confirmation to theory, RAP XY8 is robust to amplitude and frequency variation and can be especially useful for sensing in systems with large inhomogeneous broadening and amplitude inhomogeneity.

Finally, we analyzed the sensitivity to an AC field, defined as \(\eta=\frac{\sigma}{\partial S/\partial B}\sqrt{T}\) \cite{PhamPRB2012}, where $\sigma$ is the standard deviation of the single point fluorescence data in the experiment, \(\partial S/\partial B\) is the maximal slope in the curve of fluorescence signal vs. magnetic field, and \(T\) is the time for a single measurement. Figure \ref{fig:All_sensitivity} shows the robust sensitivity, i.e., lower $\eta$, of the RAP-XY8 as a function of the number of pulses when extracted indirectly from the data, in all three experiments. The difference with standard XY8 is smallest in the High Rabi case, as expected from theory, while the improvement with RAP-XY8 is significant in the presence of amplitude inhomogeneity and low Rabi frequency (inhomogeneous broadening).



\end{document}